\documentclass[aps,preprint,showpacs,superscriptaddress,groupedaddress,floatfix]{revtex4} 

\usepackage{amssymb}
\usepackage{amsmath}
\usepackage{amssymb}
\usepackage{graphicx}
\usepackage{subfigure}
\usepackage{color}
\usepackage{mathrsfs}
\usepackage[dvipsnames]{xcolor}
\usepackage[papersize={8.5in,11in}]{geometry}

\usepackage[breaklinks=true,colorlinks=true]{hyperref}
\hypersetup{colorlinks=true,citecolor=blue,linkcolor=blue,urlcolor=blue}

\usepackage[utf8]{inputenc}

\begin{document}

\title{ Kinks scattering in deformed $\tilde{\varphi}^{(6)}$ model}

\author{Aliakbar Moradi Marjaneh}
\email{moradimarjaneh@gmail.com}
\author{Azam Ghaani}
\author{Kurosh Javidan}

\affiliation{Department of Physics,  Faculty of Sciences, Ferdowsi University of Mashhad, Mashhad, Iran}

\begin{abstract}

The deformed $\tilde{\varphi}^{(6)}$ model is introduced based on the $\varphi^4$ model using a deformation functional $f[\varphi]$ including a free parameter $a$. The kink solutions in different sectors and their internal modes are obtained as functions of the deformation parameter and their characteristics are evaluated as well. It is shown that the kinks of the deformed model inherit some of their dynamical properties (like internal modes) from the standard $\varphi^4$ potential and some of their characteristics from the $\varphi^6$ model. The dynamics of kink-antikink (antikink-kink) scattering is investigated in different sectors with various kink initial conditions as well as different values of deformation parameter. According to the kinks' initial velocity, colliding kinks may be bound together or scatter from each other after the interaction. These two situations are distinguished by the critical velocity, which itself depends on the deformation parameter of the model. Due to the difference in the rest mass of kink solutions related to different sectors, interesting and sometimes rare phenomena are observed during the kink scattering and their interactions.

\end{abstract}



\maketitle

\section{Introduction}
\label{sec:introduction}
Describing various phenomena by simple mathematical equations has always been one of the goals of physicists. They usually solve the problem in its simplest form and then extend the model by adding needed realistic features to get closer to its true form. For example, nonlinear field theories are employed to describe some phenomena in high energy physics \cite{Vilenkin.book.2000,Manton.book.2004,Vachaspati.book.2006}, condensed matter \cite{Rajaraman.book.1982,Bishop.PhysD.1980,Ferro1,Ferro2,Ferro3}, and  optics \cite{Optics1,Optics2,Ozisik.Optik.2022,Ozdemir.PhysicaScripta.2023,Ozisik.Optik.2023}, etc., but scientists first examine the sine-Gordon model \cite{Rajaraman.book.1982,Moradi.EPJB.2018} as the simplest case, which is integrable with localized solutions in the form of one, two, or three kink solutions in the $(1+1)$ space-time dimensions.
Next, they delved into the study of the non-integrable $\varphi^4$ model. For instance, the authors explored the model's internal structure in references \cite{Campbell.PhisicaD.1983,Anninos.PRD.1991}. Reference \cite{Kevrekidis.book.2019} offers a comprehensive overview of various aspects of this model, while others have examined dynamical behaviors of obtained solutions from different perspectives \cite{Moradi.CNSNS.2017, Askari.CSF.2020, Mohammadi.CNSNS.2021, Pereira.JoPA.2021, Mohammadi.CSF.2022, Takyi.PRD.2023}. Notably, reference \cite{Moradi.CNSNS.2017} focuses on multi-solution collisions, and reference \cite{Askari.CSF.2020} investigates scenarios with high orders of discontinuity. Researchers have also discussed the double sine-Gordon model \cite{Peyrard.PhysD.1983.msG, Peyravi.EPJB.2009, Peyravi.EPJB.2010} and different forms of modified sine-Gordon \cite{Campbell.dsG.1986, Gani.EPJC.2018, Belendryasova.JPCS.2019, Gani.EPJC.2019} field theories as periodic models with inner structure. The $\varphi^6$ model \cite{lohe, dorey, GaKuLi, Weigel.JPCS.2014, Weigel.PRD.2016, Moradi.JHEP.2017, Gani.JOP.2020, Saadatmand.EPJB.2022, Adam.PRD.2022, Saadatmand.2023}, $\varphi^8$, and higher-order \cite{Khare.PRE.2014, Christov.PRL.2019, Khare.JPA.2019, Gani.PRD.2020, Blinov.JOP.2020, Gani.EPJC.2021, khare2021explicit} potentials are also studied extensively. These models are very practical and attractive because of their symmetric and asymmetric solutions with different masses and complex internal modes at various topological sectors. In addition, the dynamics of kink solutions of non-linear models in higher space-time (beyond the 1+1 ) dimensions have also been investigated with some new analytical or numerical methods; for example please see \cite{Mohammadi.PhysicaScripta.2020,Ozdemir.Universe.2022}.

We can look at the field $\varphi$ as the macroscopic wave function describing the phase of the system or the order parameter of the model. Crystal structures \cite{1, 2}, ferroelastic and ferroelectric fields \cite{3}, stress-induced polarization \cite {4,5} in condensed matter, massless mesons fields in particle physics \cite {lohe}, cosmological transitions in the early universe models \cite{7}, are some well-known and famous examples of such interpretations. In this context, classical $\varphi^4$ field theory successfully describes the continuous phase change as a second-order phase transition. It is due to the symmetric double well of the $V(\varphi)$ in $\varphi^4$ models. Discontinuous (first-order) phase transitions, or a certain sequence of first and second-phase transitions can not be explained by the standard $\varphi^4$ theories. Such situations can be modeled by asymmetric $\varphi^4$ models constructed by applying suitable deformations on the standard $\varphi^4$ theory and/or using higher order field theories \cite{Gani.PRD.2020, Gani.EPJC.2021}.  

Including space-dependent terms to the Lagrangian (i.e., to the equation of motion) or increasing the order of the model, provides more complex solutions with richer dynamics that may present a more realistic description of the physical phenomena. It is clear that obtaining analytical solutions for complicated nonlinear models is extremely difficult, so people analyze the behavior of such models using numerical computations and simulations. In some special cases, one can start from a certain model (like the Klein-Gordon equation) with well-known analytical solutions, and then try to transform the model into a more realistic situation, using a suitable deformation function. Thus, the solutions of the new model are also obtainable with the help of an associate deformation function \cite{Bazeia.PRD.2002,Bazeia.PRD.2004,Bazeia:2005hu,Bazeia.PRD.2006,Bazeia.EPJC.2018,Moradi.CSF.2022}.\

Motivated by this issue, we have studied the dynamics of a new deformed model applied to the $\varphi^4$ field theory. Through this procedure, the standard $\varphi^4$ model transforms into a deformed $\tilde{\varphi}^{(6)}$ field theory. Thus, we will find new kink solutions of a deformed $\tilde{\varphi}^{(6)}$ potential using standard kinks of the $\varphi^4$ model. Considering the very different properties and dynamics of $\varphi^4$ and $\varphi^6$ models, we should expect new solutions that inherit some of their properties from the $\varphi^4$ field theory and some others from the standard $\varphi^6$ model. 

The structure of the paper is as follows: In the next section, we will introduce the applied deformation functional and its characteristics. Section \ref{sec:modifiedphi6model} is dedicated to the construction of the deformed model $\tilde{\varphi}^{(6)}$ by applying the introduced deformation to the $\varphi^4$ potential. The kink (antikink) solutions of the deformed model in different sectors, the rest mass of different kinks, and the internal modes related to each solution will be examined in this chapter, too. Kink-antikink (antikink-kink) collisions in different sectors with several initial conditions will be presented in section \ref{sec:Multi-kink scattering}, and finally section \ref{sec:conclusion} is devoted to summaries and conclusions. 

\section{General statement, The deformation functional}
\label{sec:deformationfunction}

The dynamic of a real scalar field model in $(1+1)$ space-time  dimensions is described by the Lagrangian density:
\begin{eqnarray}\label{eq:lagrangian}
\mathcal{L}=\frac{1}{2}\left(\frac{\partial \varphi}{\partial t}\right)^2-\frac{1}{2}\left(\frac{\partial \varphi}{\partial x}\right)^2-V(\varphi).
\end{eqnarray}
This Lagrangian yields the Klein-Gordon equation of motion:
\begin{equation}\label{eq:EOM}
\varphi_{tt} - \varphi_{xx} + \frac{dV(\varphi)}{d\varphi}=0.
\end{equation}
In the static case, $\varphi=\varphi(x)$, Eq.~(\ref{eq:EOM}) can be reduced to the following Bogomolny equation:
\begin{equation}\label{eq:bps}
\frac{d\varphi}{dx} = \pm\sqrt{2V(\varphi)}.
\end{equation}
The above equation provides two solutions for positive and negative signs. According to the boundary conditions on the potential function, the solution with positive sign is called kink solution ($K$) and the solution with negative sign is called antikink ($\bar{K}$). Static kink ($K$) and antikink ($\bar{K}$) solutions can be found analytically (or numerically) by solving the first-order ordinary differential Eq.~(\ref{eq:bps}). 

 Deformed soliton solutions have received much attentions because of their new physics \cite {Ceschin.JHEP.2021, Rutkevich.JHEP.2024}. Such deformations come from some limitations in practical fabrications and medium defects \cite {Chekmazov.SciReps.2024}. Because of nonlinear nature of these models, small changes in model parameters, create unexpected results. Indeed, we have to find new localized solutions and reexamine their characteristics, using standard methods applied in nonlinear dynamics. Here we study a nonautonomous deformed procedure appears in practical situations, like q-deformation models. Such change in the field model introduce new exact solutions with different dynamics.

By applying a deformation functional, $f[\varphi]$, on the potential $V(\varphi)$ \cite{Bazeia.PRD.2002, Bazeia.PRD.2004, Bazeia:2005hu, Bazeia.PRD.2006, Bazeia.EPJC.2018, Moradi.CSF.2022}, we arrive at a deformed potential $\Tilde{V}(\varphi)$:

\begin{equation}\label{eq:deformedpotential}
\Tilde{V}(\varphi)=\frac{V[\varphi\to f(\varphi)]}{[f^{\prime}(\varphi)]^2},
\end{equation}
with new static kinks which can be calculated as: 
\begin{equation}\label{eq:deformedkinks}
\Tilde{\varphi}_K(x)=f^{-1}[\varphi_K(x)].
\end{equation}
Time-dependent (moving) kink solution is derived by applying a Lorentz boost to the static solution. The classical energy (mass) of kink or antikink is calculated by integrating Hamiltonian density over space:

 \begin{eqnarray}\label{eq:energy}
E(t) &=& \int_{-\infty}^{+\infty} \left(\frac{1}{2}\left(\frac{\partial \varphi}{\partial t}\right)^2+\frac{1}{2}\left(\frac{\partial \varphi}{\partial x}\right)^2+V(\varphi)\right) dx.  
 \end{eqnarray}
The rest mass of the kink (antikink) is obtained by substituting static solution $\varphi_K(x)$ or its deformed counterpart (\ref{eq:deformedkinks}) into (\ref{eq:energy}) which leads to:
 \begin{eqnarray}\label{eq:mass}
M_0 &=& \int_{-\infty}^{+\infty} \left(\frac{\partial \varphi_K}{\partial x}\right)^2 dx.  
 \end{eqnarray}

To check for the presence of possible internal mode(s), one can add the small perturbation to the static kink $\varphi(x)$, as $\varphi(x,t)=\varphi(x)+\epsilon \eta(x) \cos (\omega t)$. Perturbed kink is substituted back into (\ref{eq:EOM}) and linear terms of the obtained equation are collected up to the order $O(\epsilon)$. The resulting linearised field equation is: 

\begin{equation}\label{eq:schrodingerlike}
-\frac{\partial^2 \eta}{\partial \varphi^2}+U\eta=\omega^2\eta.
\end{equation}
The function $U=U(x)$ is the kink stability potential and it is obtained from the second derivative of the potential $V(\varphi)$ ( $\Tilde{V}(\varphi)$ ),  respect to the $\varphi$ ($\Tilde{\varphi}$):
\begin{eqnarray}\label{eq:stabilitypotentials}
U(x)=\frac{d^2V(\varphi)}{d\varphi^2}|_{\varphi=\varphi(x)}.
\end{eqnarray}
The existence of internal mode and characters of kink excitation spectra depend on the behavior of the possible solutions of the above Schr\"odinger-like equation.

Now we have all the necessary preparations to introduce the new model. In the next section, we will present the deformation procedure and its application on a specific $\varphi^4$ model.
\section{modified \texorpdfstring{$\tilde{\varphi}^{(6)}$ }{pdfbookmark} model}
\label{sec:modifiedphi6model}

\begin{figure*}[!ht]
\begin{center}
  \centering
    \subfigure[]{\includegraphics[width=0.45
 \textwidth]{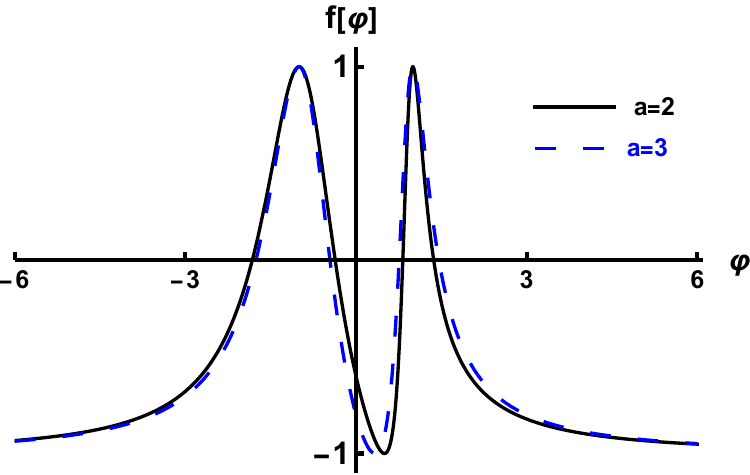}\label{fig:f}}
  \subfigure[]{\includegraphics[width=0.45
 \textwidth]{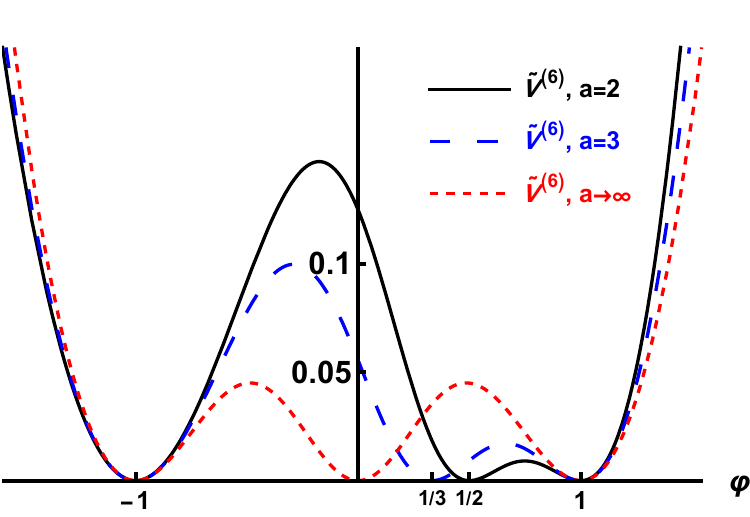}\label{fig:modifiedphi6potentials}}
\\
  \subfigure[]{\includegraphics[width=0.45
 \textwidth]{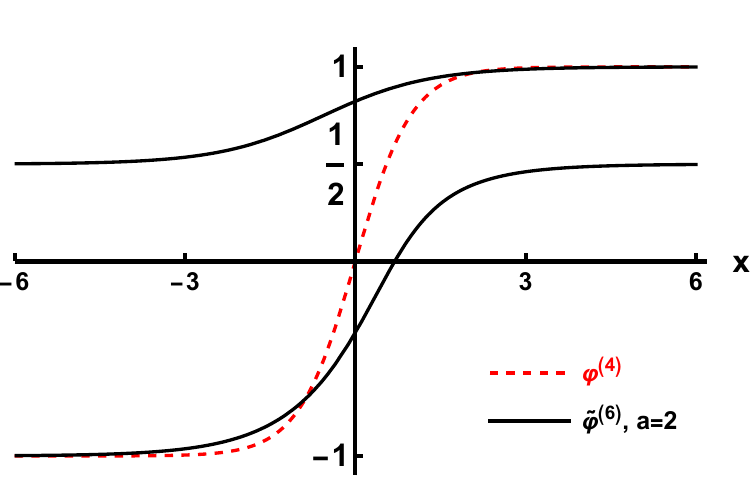}\label{fig:modifiedphi6kinksa2}}
  \subfigure[]{\includegraphics[width=0.45
 \textwidth]{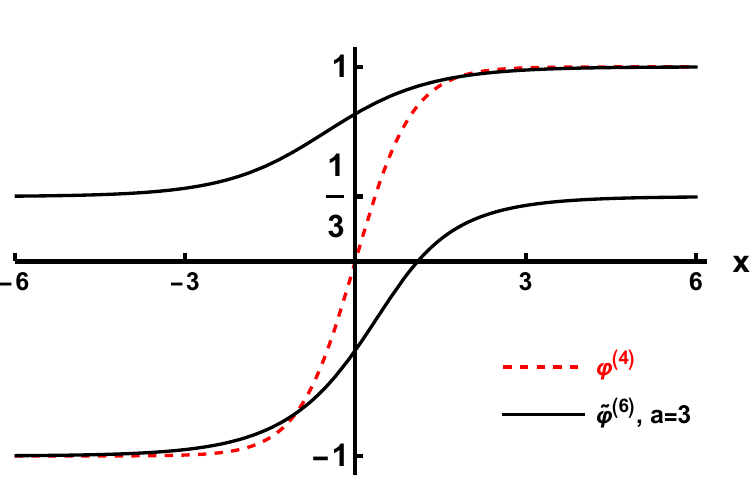}\label{fig:modifiedphi6kinksa3}}
\\
  \caption{(a) Deformation function, Eq.~(\ref{eq:f1}) for $c=0$, (b) potential, Eq.~(\ref{eq:modifiedphi6potential}), kink solution of modified $\tilde{\varphi}^{(6)}$ model with (c) $a=2.0$ and (d) $a=3.0$. }
  \label{fig:modifiedphi6}
\end{center}
\end{figure*}

The deformation functional $f[\varphi]$ is defined as (Fig.~\ref{fig:f}):

\begin{equation}\label{eq:f1}
f[\varphi]=\frac{\left(\frac{1}{a}-\varphi \right)^2-e^{2 c} \left(\varphi ^2-1\right)^2}{\left(\frac{1}{a}-\varphi \right)^2+e^{2 c} \left(\varphi ^2-1\right)^2}.
\end{equation}
where $a$ and $c$ are model-adjusting parameters.
We have deformed standard $\varphi^4$ potential $V^{(4)}=\frac{1}{2}(1-\varphi^2)^2$ using (\ref {eq:f1}) through (\ref {eq:deformedpotential}) with $c=0$, which results a deformed $\tilde{\varphi}^{6}$ field theory as follows: 
\begin{eqnarray}\label{eq:modifiedphi6potential}
\tilde{V}^{(6)}&=&\frac{1}{2}\frac{\left(\varphi ^2-1\right)^2 \left(\varphi -\frac{1}{a}\right)^2}{ \left(\varphi ^2-\frac{2}{a}\varphi+1\right)^2}. 
\end{eqnarray}
The new potential has three minima at $\pm1$ and $\frac{1}{a}$. Thus the model contains two sectors $(-1,\frac{1}{a})$ and $(\frac{1}{a},1)$. So if we consider real values greater than one for $a$, (i.e. $a>1$) this parameter controls the position of a vacuum between the other two minima.

Figs.~\ref{fig:f} and \ref{fig:modifiedphi6potentials} demonstrate deformation functional $f\left[\varphi\right]$ and deformed $\tilde{V}^{(6)}$ potential as functions of the field $\varphi$. These figures clearly show that the results of such deformation are not symmetrical with respect to the transformation $\varphi \leftrightarrow -\varphi$. For this reason, the characteristics and dynamics of kinks in different sectors are not the same. This means that the position of the central minimum (which is determined by the value of $a$) has a very decisive effect on the characteristics of the model and its general behavior. Due to the critical dependence of the characteristics of sectors on the parameter $a$, the dynamics of kink solutions in each sector are very sensitive to the value of parameter $a$. Thus, the behavior of the kink solution in the deformed model will be very different from the dynamics of kinks created in the well-known model $V^{(6)}=\frac{1}{2}\varphi^2(1-\varphi^2)^2$. In the limit $a\rightarrow \infty$, the deformed potential $\tilde{V}^{(6)}$ changes into the standard potential $V^{(6)}$, as one can find from (\ref{eq:modifiedphi6potential}) (also see Fig.~\ref{fig:modifiedphi6potentials}). 

 Kinks of the standard model $V^{(6)}$ are obtained by solving the Eq.~(\ref{eq:bps}). Then, we have derived all explicit asymmetric kink solutions of the deformed model (\ref{eq:modifiedphi6potential}) by applying a transformation (\ref{eq:deformedkinks}) to the kink (antikink) solutions of the standard $V^{(4)}$ modelas follows:
\begin{eqnarray}\label{eq:modifiedphi6kinks}
\begin{cases}
\Tilde{\varphi}^{(6)}(-1,\frac{1}{a}) = \frac{1}{2} \left(e^x-\sqrt{-\frac{4 e^x}{a}+e^{2 x}+4}\right)\\ 
\Tilde{\varphi}^{(6)}(\frac{1}{a},1)=-\frac{1}{2} e^{-x} \left(1-\sqrt{\frac{4 e^x}{a}+4 e^{2 x}+1}\right),
\end{cases}
\end{eqnarray}

Figs.~\ref{fig:modifiedphi6kinksa2} and \ref{fig:modifiedphi6kinksa3} present modified kink solutions as functions of  $x$ for different values of parameter $a=2$ and $a=3$, respectively. Eq.~(\ref{eq:modifiedphi6kinks}) as well as Figs.~\ref{fig:modifiedphi6kinksa2} and \ref{fig:modifiedphi6kinksa3} show that the kink solutions are not symmetrical in any of the sectors. For simplicity, we will call sector $\Tilde{\varphi}^{(6)}(\frac{1}{a},1)$ the "up" sector and sector $\Tilde{\varphi}^{(6)}(-1,\frac{1}{a})$ the "down" sector from here on.

The energy of moving (standard or modified) kinks  and their rest mass is calculated using (\ref{eq:energy}) and (\ref{eq:mass}), respectively. The rest mass of modified kinks $\tilde{\varphi}$ related to the different sectors are obtained as:
\begin{eqnarray}\label{eq:modifiedphi6mass}
\begin{cases}
\tilde{M}^{(6)}_{(\frac{1}{a},1)}=\frac{2 \sqrt{a^2-1}}{a^2}\tan ^{-1}\left(\frac{a-1}{\sqrt{a^2-1}}\right)-\frac{a-1}{2 a^2} \left(a+3-2 (a+1) \ln \left(\frac{2 a}{a+1}\right)\right),
\\
\tilde{M}^{(6)}_{(-1,\frac{1}{a})}=-\frac{2 \sqrt{a^2-1}}{a^2}\tan ^{-1}\left(\frac{a+1}{\sqrt{a^2-1}}\right)+\frac{(a+1)}{2 a^2}\left(-a+3+2 (a-1) \ln \left(\frac{2 a}{a-1}\right)\right).
\end{cases}
\end{eqnarray}

\begin{figure*}[!ht]     
\begin{center}
  \centering
    \subfigure[ ]{\includegraphics[width=0.45
 \textwidth]{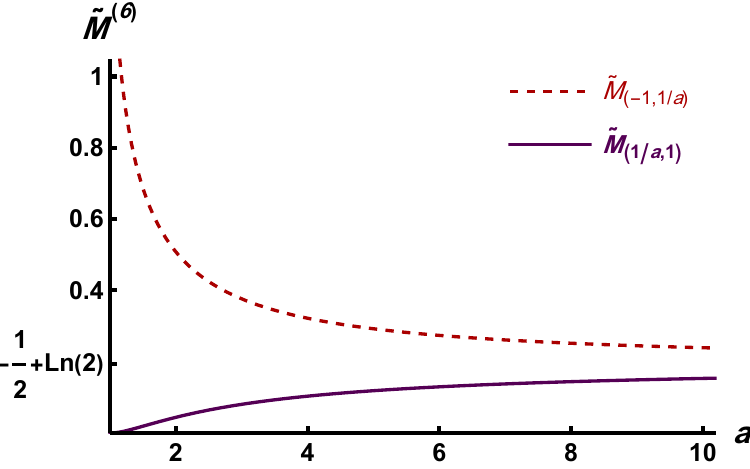}\label{fig:modifiedphi6mass}}
  \subfigure[ ]{\includegraphics[width=0.45
 \textwidth]{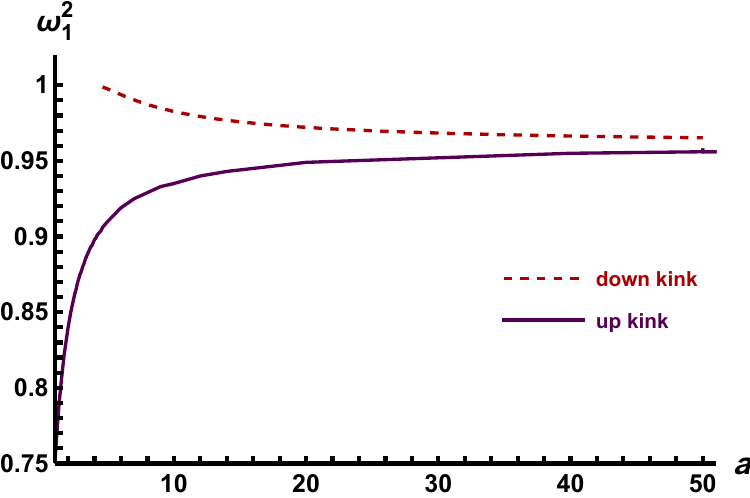}\label{fig:modifiedphi6modes}}
\\
    \subfigure[ ]{\includegraphics[width=0.32
 \textwidth]{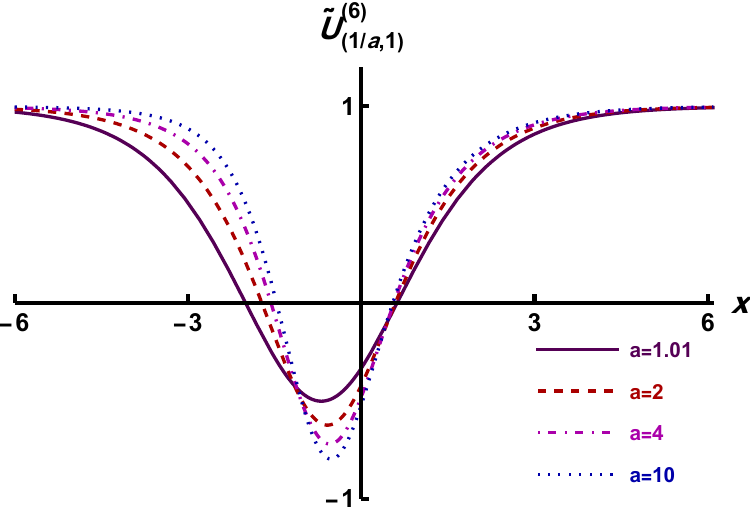}\label{fig:modifiedphi6qmpup}}
  \subfigure[ ]{\includegraphics[width=0.32
 \textwidth]{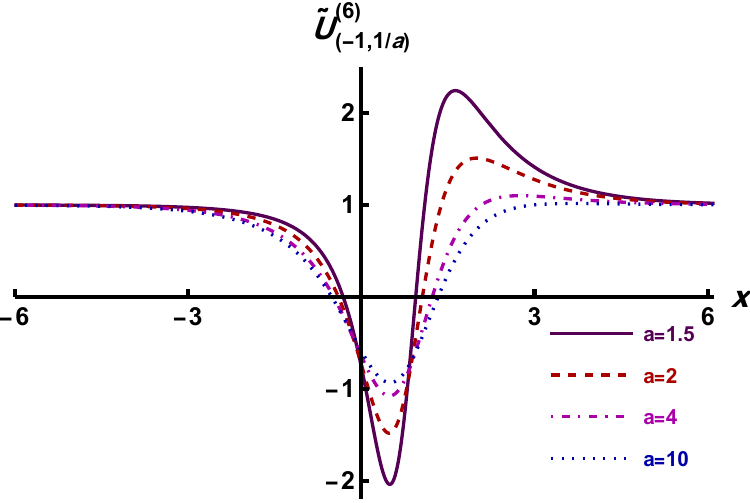}\label{fig:modifiedphi6qmpdawn}}
  \subfigure[ ]{\includegraphics[width=0.32
 \textwidth]{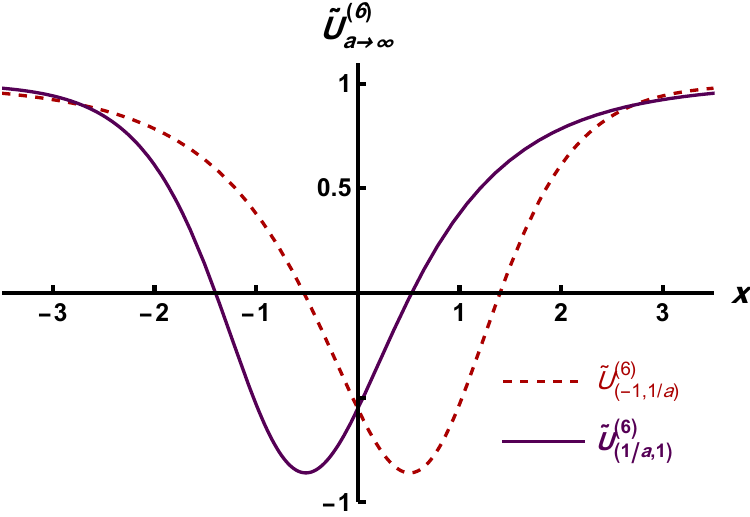}\label{fig:modifiedphi6qmpInf}}
\\
  \caption{ The modified $\tilde{\varphi}^{(6)}$  (a) kink's mass, Eq.~(\ref{eq:modifiedphi6mass}), (b) kink's internal modes as functions of parameter $a$,  
  (c), (d) quantum mechanical potential for $\Tilde{\varphi}^{(6)}(\frac{1}{a},1)$ and $\Tilde{\varphi}^{(6)}(-1,\frac{1}{a})$, respectively (Eq.~\ref{eq:modifiedphi6kinks}), as a function of $x$ and (e) quantum mechanical potential for $a \to \infty$ .}
  \label{fig:modifiedphi6massmodesqmp}
\end{center}
\end{figure*}

The rest mass of both kinks become $M_0=\left(\ln{(2)}-\frac{1}{2} \right)$ at $a \to \infty$ as Fig.~\ref{fig:modifiedphi6mass} shows. On the other hand, at the limit $a \rightarrow1$, the kink rest mass in one of the sectors tends to zero, while the rest mass of the kink in the other sector becomes very high. The rest mass difference between the kinks of two sectors in a model can produce complicated behavior and interesting dynamics.

To find the internal mode(s) of the deformed model, we need to solve Eq.~(\ref{eq:schrodingerlike}) with kink stability potential Eq.~(\ref{eq:stabilitypotentials}) for both sectors separately. These potentials have been plotted in Figs.~\ref{fig:modifiedphi6qmpup} and \ref{fig:modifiedphi6qmpdawn}. It is clear that $\tilde{V}^{(6)}$ model (\ref{eq:modifiedphi6potential}) is not integrable. Internal modes of some nonintegrable models are responsible for periodic changes in the speed, amplitude, or spatial profile of kink solutions. Such oscillations are absent in the dynamical behavior of integrable systems. It has been shown that the $\varphi^4$ model supports an internal mode \cite{Campbell.PhisicaD.1983,Anninos.PRD.1991,Kevrekidis.book.2019,Moradi.CNSNS.2017}. Indeed, the multi-bounce resonance in the scattering of the $\varphi^4$ localized solutions has been generally explained based on energy exchange between a kink solution and its internal mode that oscillates around the kink.  However, the standard $\varphi^6$ field theory is not integrable, the single kink (antikink) of this model does not have internal modes \cite{lohe,dorey,GaKuLi,Weigel.PRD.2016}. Thus, we do not expect to find multi-bounce behavior due to kink-internal mode interaction in different forms of kink-antikink or antikink-kink scattering. Despite this issue, antikink-kink scatterings with specific initial conditions, also exhibit resonant interactions \cite{dorey}. This means that multi-bounce scattering can occur due to a kinetic energy exchange between the kinks and bound state collective potential well produced by $\bar{K}K$ pair.\

We have constructed the modified $\tilde{V}^{(6)}$ model based on the standard $\varphi^ 4$ potential. Although the $\varphi^6$ model does not have internal modes, the new deformed field theory has inherited the existence and number of internal modes from the $\varphi^4$ model. The outline of the quantum mechanical potential $\tilde{U}^{(6)}_{(\frac{1}{a}, 1)}$ is in accordance with quantum mechanical potential of the $\varphi^4$ model, while $\tilde{U}^{(6)}_{(-1,\frac{1}{a})}$ is very similar to derived related function from $\varphi^6$ potential. Therefore, it can be assumed that the kink solution of the up sector generally contains an internal mode, but the kink of the down sector does not have an internal mode, at least for some values of the deformation parameter $a$. Fig.~\ref{fig:modifiedphi6modes} demonstrates eigenfrequencies of Eq.~(\ref{eq:schrodingerlike}) for up and down sectors as functions of parameter $a$. The kink solution of the up sector has an internal mode for all values of the deformation parameter. On the other hand, the kink solution of the down sector does not have an internal mode for values of the deformation parameter close to one (maximum deformation), but one internal mode is obtained for $a \ge 4.6$. It is an interesting outcome.  As illustrated in Figs.~\ref{fig:modifiedphi6qmpup} and \ref{fig:modifiedphi6qmpdawn}, increasing parameter $a$, 
 results in a deeper quantum mechanical potential well for the up-sector kink, while the well for the down-sector kink becomes shallower. When parameter $a$ approaches infinity, $a\to \infty$, the depths of both wells become nearly identical, as shown in Fig.~\ref{fig:modifiedphi6qmpInf}. Consequently, the kinks internal mode for the up-sector and down-sector converge to the same value. Indeed, the overall results from Figs.~\ref{fig:modifiedphi6qmpup}, \ref{fig:modifiedphi6qmpdawn} and \ref{fig:modifiedphi6qmpInf} are consistent with what we have expected.

\section{two kinks scattering}
\label{sec:Multi-kink scattering}

Bound states are formed at low energies of the $\bar{K}K$ or $K\bar{K}$ scatterings. Two different mechanisms can be activated to form a possible final bound state. One is the interaction of the kinks with the internal mode and the other is the interaction between the kinks and bound state collective potential well, which happens only in the $\bar{K}K$ \cite{dorey}. The kink solution of the up sector has an internal mode for all values of parameter $a$. During the $\bar{K}K$ interaction, both mechanisms are activated, while in the $K\bar{K}$ system, only the internal mode is involved in the formation of the bound state. By increasing the deformation parameter $a$, the mass of the kink in the up sector increases (see Fig.~\ref{fig:modifiedphi6mass}). Assuming a slight change in binding energy, increasing the mass decreases the critical velocity. But the depth of collective potential well in $\bar{K}K$ increases due to the increase of the kink amplitude (compare kink amplitudes in Figs.~\ref{fig:modifiedphi6kinksa2} and \ref{fig:modifiedphi6kinksa3}). For this reason, bound states are also established by taking larger values of $a$ at higher energies (higher critical velocity).

\begin{figure*}[!ht]
\begin{center}
  \centering
    \subfigure[ ]{\includegraphics[width=0.32
 \textwidth]{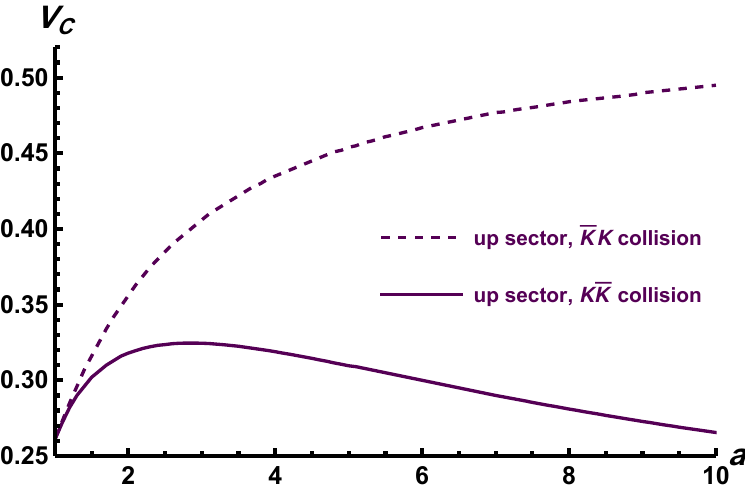}\label{fig:CriticalVelocityUp}}
    \subfigure[ ]{\includegraphics[width=0.32
 \textwidth]{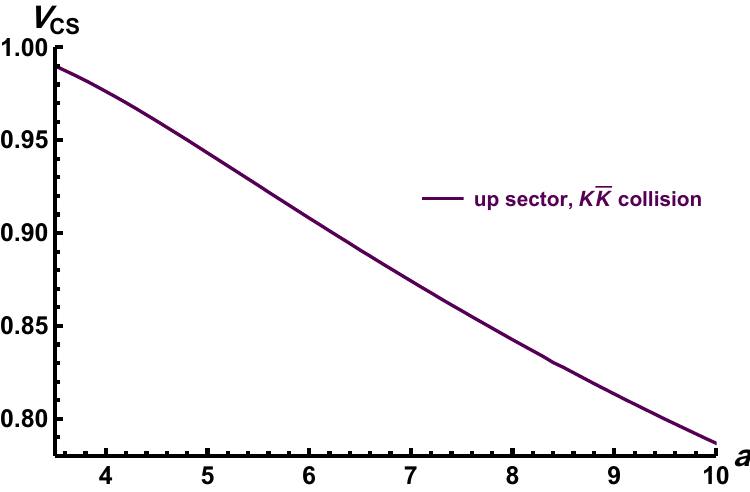}\label{fig:ChangingSectorVelocityUp}}
  \subfigure[ ]{\includegraphics[width=0.32
 \textwidth]{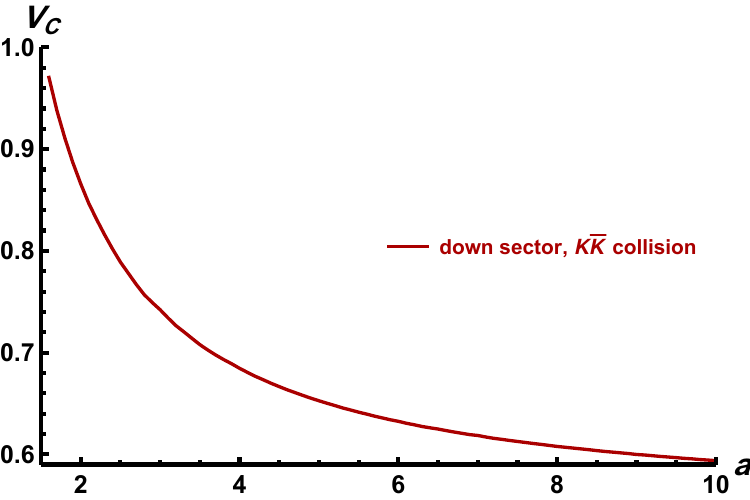}\label{fig:CriticalVelocityDown}}
\\
  \caption{ Scattering of modified $\tilde{\varphi}^{(6)}$ solutions at different sectors with initial conditions and numerical parameters $2X_0=40, Nat=5000, h=0.05, \delta t=0.005$,  (a) critical velocity as a function of $a$ for two kinks collision in up sector, (b) changing sector velocity as a function of $a$ for kink-antikink collision in up sector and (c) critical velocity as a function of $a$ for kink-antikink collision in down sector. }
  \label{fig:CriticalVelocity}
\end{center}
\end{figure*}

 Fig.~\ref {fig:modifiedphi6mass} shows that the static mass of the kink in the up sector is smaller than the kink mass in the down sector but the mass difference becomes smaller when the parameter $a$ increases. 

Process of $K\bar{K}$ ($\bar{K}K$) Interaction has been studied using the numerical solution of the nonlinear partial differential equation (\ref {eq:EOM}), with various initial conditions. The spatial derivatives in the differential equation have been discretized by the finite difference method with grid spacing $h$, while the time derivative term is approximated using the $4^{th}$ order Runge-Kutta method with time steps $\delta t<h^2$. We have used the fixed boundary conditions in our calculations. Kink and antikink have been taken well separated in the initial conditions so that there is no significant overlap between the initial kinks. In addition, we take the spatial interval of calculations big enough in a way that the moving kinks and energy radiations do not reach the boundary throughout the time evolution. To ensure that the results do not depend on the numerical calculation parameters (i.e. $h$, $\delta t$ , …) and to control the validity of the numerical results, we have performed numerical calculations with different values for time-space grid lengths and compared the results. Our simulations show that $h=0.05$, 
$\delta t=0.005$, total grid space length $N=5000$ and initial kink separation $2X_0=20$ are acceptable simulation parameters.

In a general overview, two distinct types of scattering are distinguished in the kink-antikink collisions. For velocities less than a certain value $V_C$, two colliding objects are temporarily or permanently bound together after encountering the radiation of a certain amount of energy. Such collision is relatively inelastic, and we examine such processes in detail further. For collisions with larger initial velocities than the $V_C$, the kinks scatter to each other in an (almost) elastic interaction. In deformed potentials, the critical velocity $V_C$ depends on the deformation parameter in addition to the model parameters. 

To find the critical velocity $V_C$, we have setup several kink collisions with different values of initial velocity. We usually start from high velocities where the $K\bar{K}$ ($\bar{K}K$) interaction is elastic. The largest initial velocity at which the bound state is formed is close to the critical speed. Now, with the required accuracy, we can inspect the nearest value of initial velocity to the boundary of changing the behavior.

Figs.~ \ref {fig:CriticalVelocityUp} and \ref {fig:CriticalVelocityDown} demonstrate critical velocity $v_C$ as functions of deformation parameter $a$ for $K\bar{K}$ and $\bar{K}K$ collisions in up and down sectors. Nevertheless, the kink-antikink collision in the up sector (of course, with sufficiently large velocity) can lead to their annihilation and the creation of kink-antikink pair(s) in the down sector, especially for larger values of parameter $a$. Fig.~\ref{fig:ChangingSectorVelocityUp} shows the minimum velocity ($V_{CS}$) required for this interaction. As stated before, the difference in the mass of the kink in the up and down sectors decreases with the increase of $a$, and as a result, the minimum velocity $V_{CS}$ also decreases. The change of sector in $K\bar{K}$ collisions, especially for the transition from a sector with a lighter mass to a sector with a heavier mass, has rarely been observed in nonlinear models. This mechanism can be used to describe some important phenomena in condensed matter, particle physics, etc.
\begin{figure*}[!ht]
\begin{center}
  \centering
      \subfigure[$a=2.0, v_i=0.3157, v_f=0.0$,  n-bounce windows ]{\includegraphics[width=0.45
 \textwidth,height=0.2 \textheight]{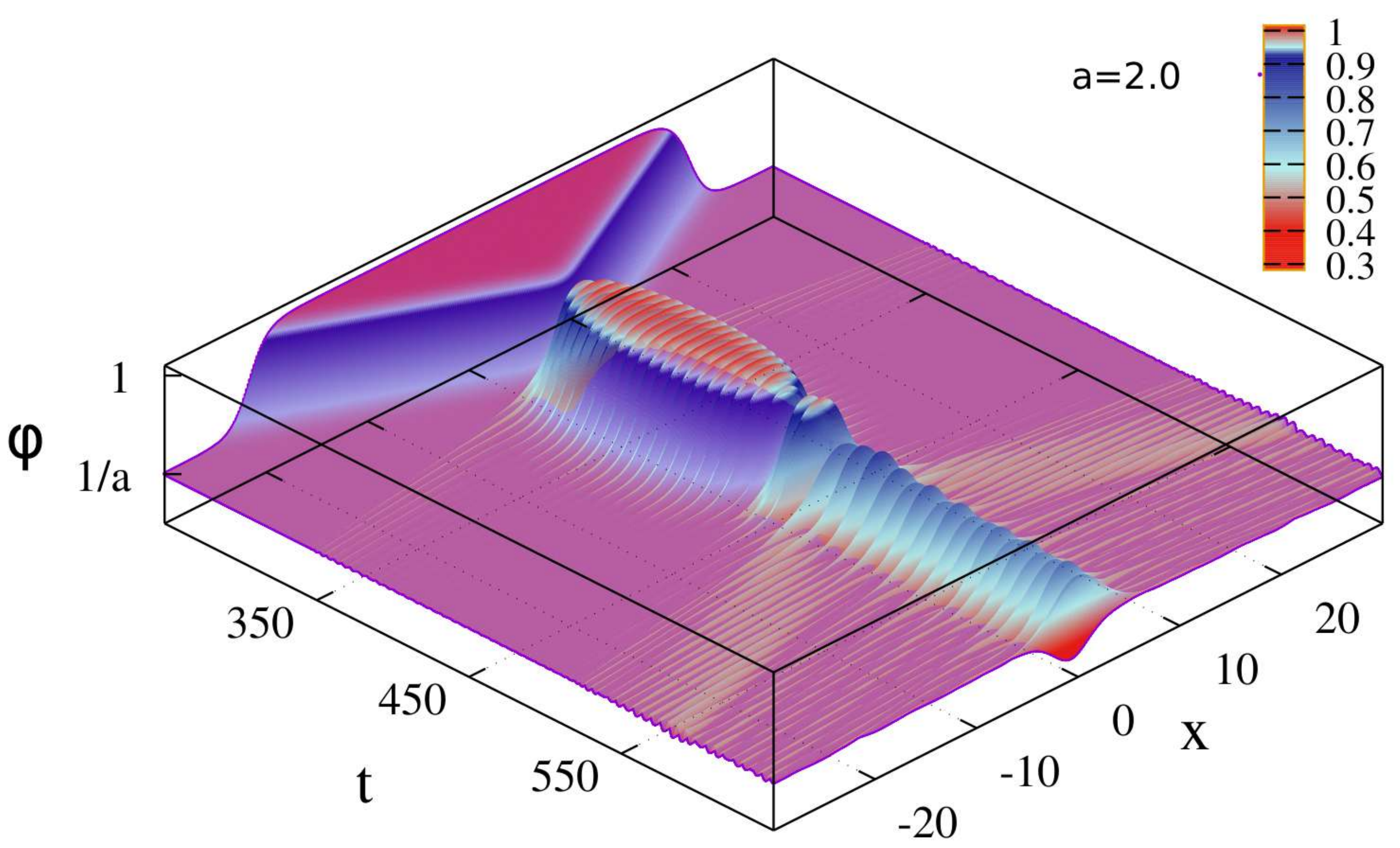}\label{fig:upKakX0100V03157vf00000a20}}
      \subfigure[$a=4.1, v_i=0.3157, v_f=0.1599$, 3-bounce windows]{\includegraphics[width=0.45
 \textwidth,height=0.2 \textheight]{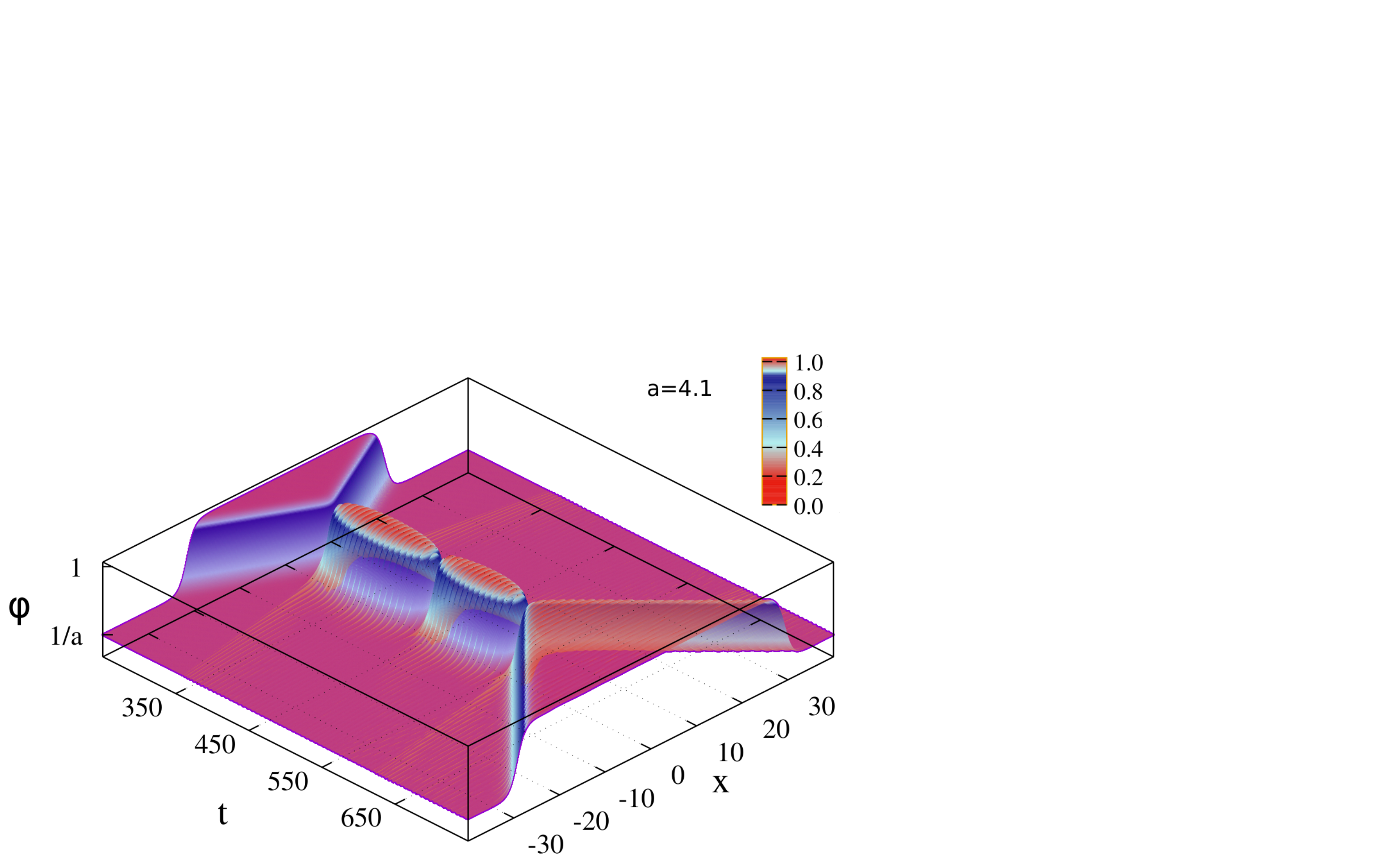}\label{fig:upKakX0100V03157vf01599a41}}
\\
      \subfigure[$a=2.0, v_i=0.3166, v_f=0.1827$, 2-bounce windows]{\includegraphics[width=0.45
 \textwidth,height=0.2 \textheight]{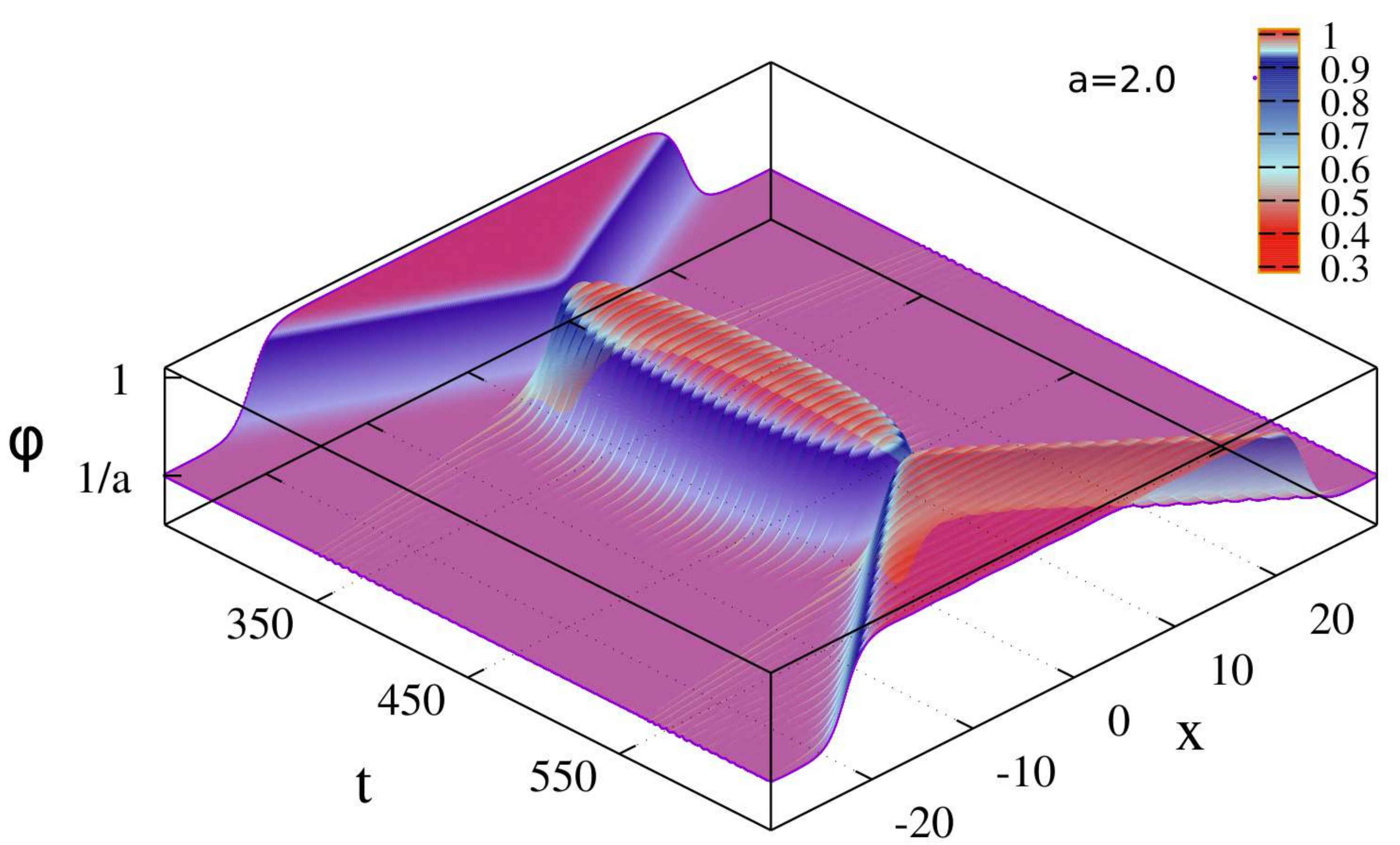}\label{fig:upKakX0100V03166vf01827a20}}
      \subfigure[$a=4.1, v_i=0.3166, v_f=0.0992$, 3-bounce windows ]{\includegraphics[width=0.45
 \textwidth,height=0.2 \textheight]{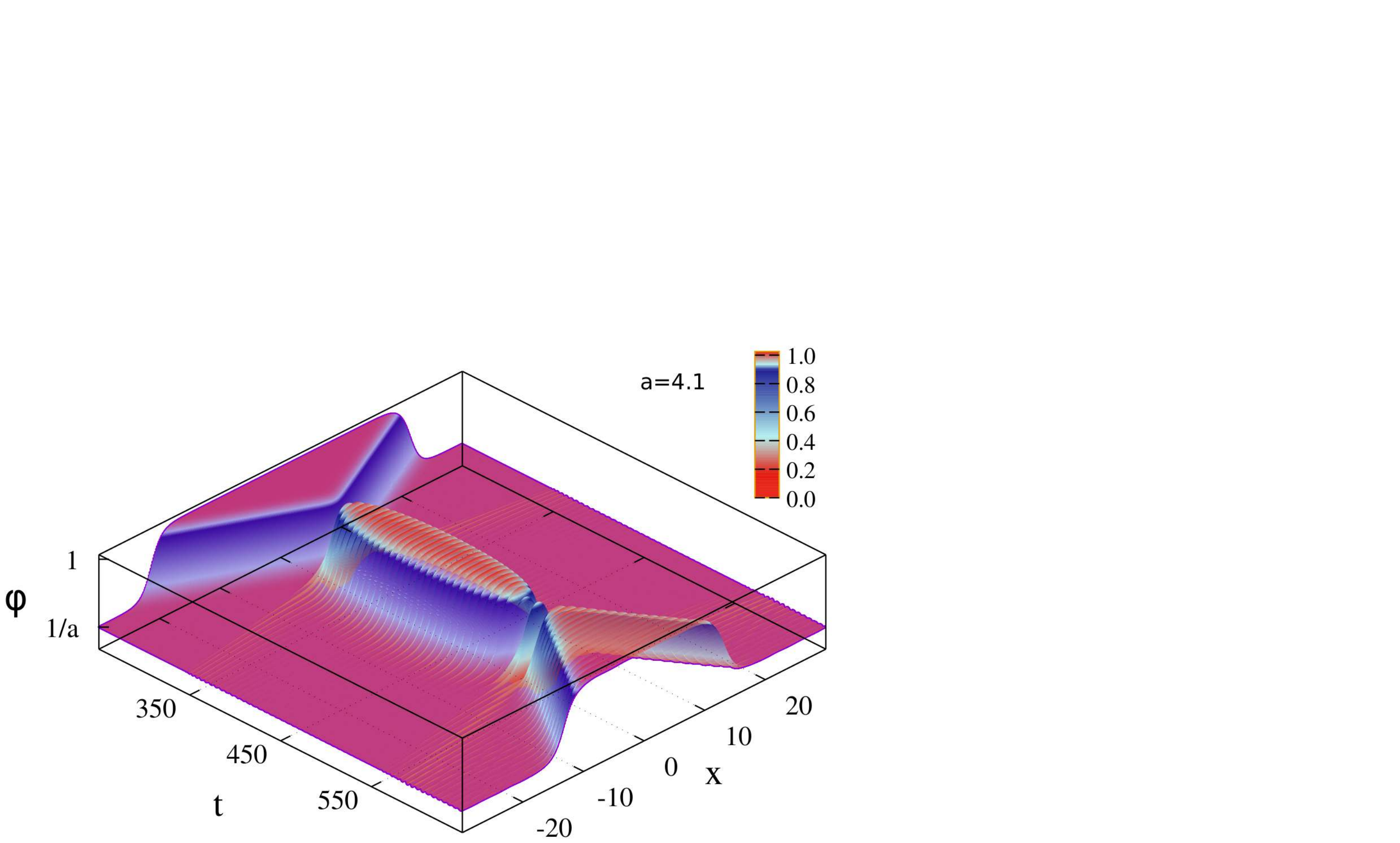}\label{fig:upKakX0100V03166vf00992a41}}
\\
      \subfigure[$a=2.0, v_i=0.9800, v_f=0.9312$, scattering for $v>v_C$ ]{\includegraphics[width=0.45
 \textwidth,height=0.2 \textheight]{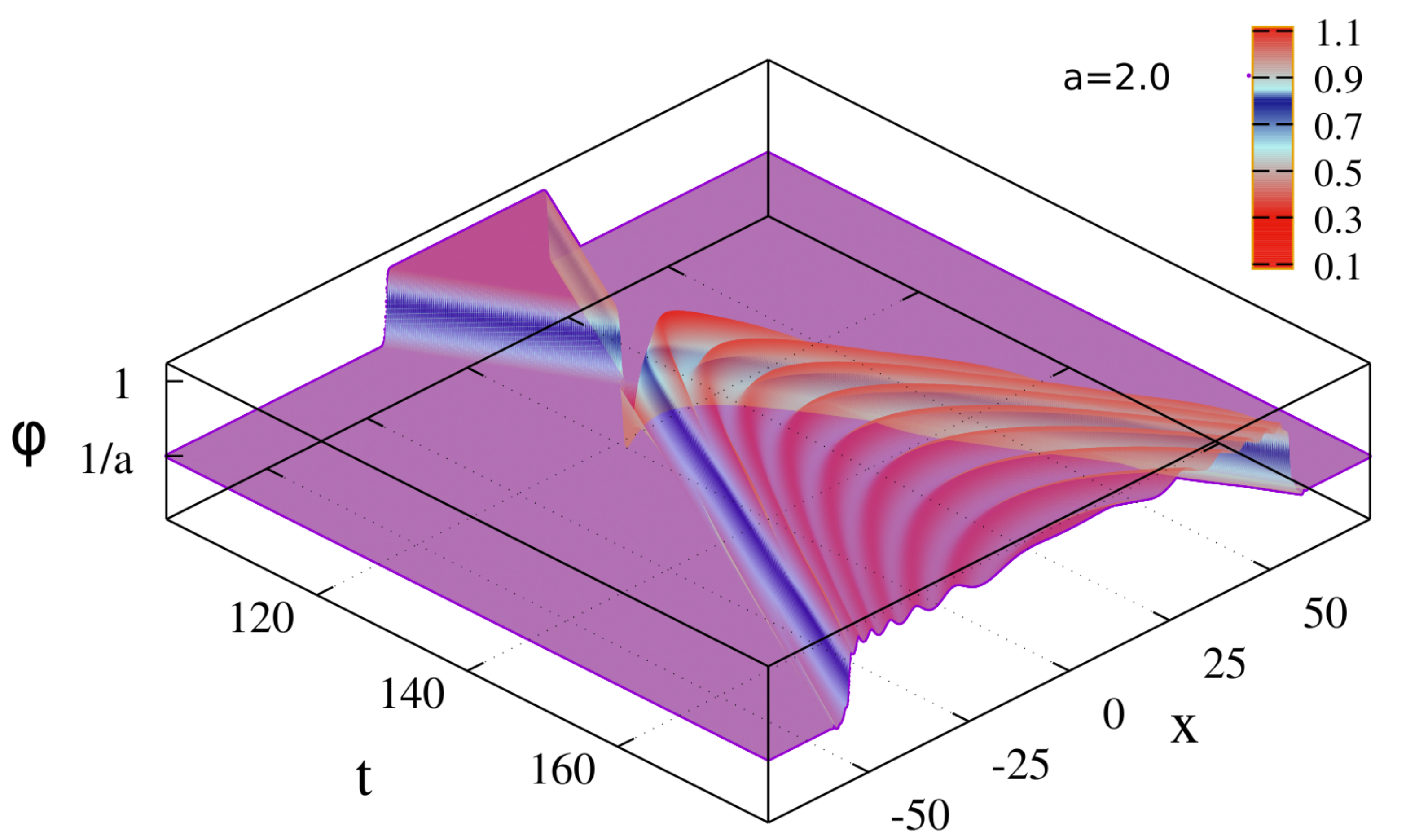}\label{fig:upKakX0100V09800Vf09312a20}}
      \subfigure[$a=4.1, v_i=0.9800, v_f=0.4828$, changing sector for $v > v_{CS}$ ]{\includegraphics[width=0.45
 \textwidth,height=0.2 \textheight]{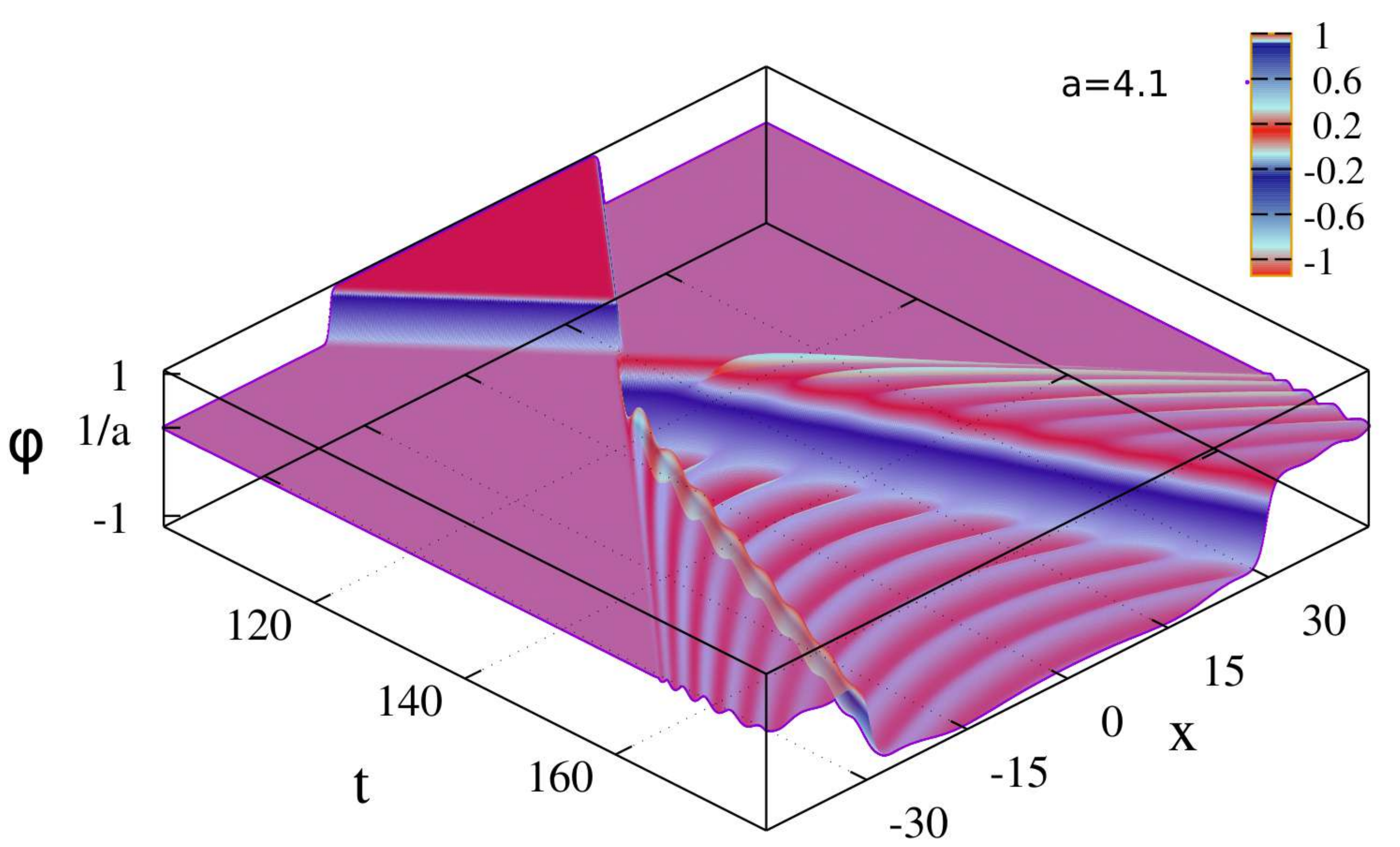}\label{fig:upKakX0100V09800Vf04828a41}}
\\
  \caption{ $V_C=0.3188$ for both $a=2.0$ and $a=4.1$. $v_i=0.3157<V_C$: (a) $a=2.0$ and (b) $a=4.1$. $v_i=0.3157<V_C$ for: (d) $a=2.0$ and (d) $a=4.1$. $v_i=0.08>V_C$ for (e) $a=2.0$ elastic scattering no sector change and (f) $a=4.1$ elastic scattering with sector change.} 
  \label{fig:upkak}
\end{center}
\end{figure*}
As the kink solution of the down sector is heavier than that in the up sector, the $K \bar{K}$ scattering in the down sector with higher velocities leads to the creation of different types of bound states. By increasing the deformation parameter $a$, the kink mass decreases, and therefore, bound states occur at lower velocities. In other words, in the down sector, the critical velocity $V_C$ decreases, as the deformation parameter increases. Fig.~\ref {fig:CriticalVelocityDown} presents this situation.

The position of multi-bounce windows in velocity space and their orders are still not clearly understood in nonlinear models. In general, there exist three-bounce resonance windows surrounding each two-bounce window in some nonlinear field theories. Similarly, four-bounce windows surround each three-bounce window, and indeed narrower and narrower windows can be found in velocity space. Figs.~\ref{fig:upkak} show the kink-antikink collision in the up sector for different values of deformation parameter $a$, ($a=2$ and $a=4.1$ which have the same critical velocity, $v_C=0.3180$, in kink-antikink collision) and different initial velocity. Figs.~\ref{fig:upKakX0100V03157vf00000a20} and \ref{fig:upKakX0100V03157vf01599a41} have been plotted for two different values of the deformation parameter ($a=2.0$ and $a=4.1$) but a same initial velocity $v_i=0.3157$. The initial velocity has been chosen lower than the critical velocity $V_C$ for both values of the deformation parameter. Therefore, we expect to see bound states for both initial conditions. In Fig.~\ref{fig:upKakX0100V03157vf00000a20}, the result is n-bounce interaction, while for the initial conditions of Fig.~\ref{fig:upKakX0100V03157vf01599a41}, we have found 3-bounce collision. With a slight change in the initial velocity, the bounce window may change or the interaction result may remain unchanged. Figs.~\ref {fig:upKakX0100V03166vf01827a20} and \ref {fig:upKakX0100V03166vf00992a41} demonstrate the $K \bar{K}$ interaction with another value for the kink initial velocity $v_i=0.3166$ but the same values for the deformation parameter as Figs.~\ref{fig:upKakX0100V03157vf00000a20} and \ref{fig:upKakX0100V03157vf01599a41} respectively. For $a=2.0$, the bounce window changes from n-bounce into 2-bounce scattering, while for $a=4.1$ we still find the same interaction in the 3-bounce window.

Figs.~\ref{fig:upKakX0100V09800Vf09312a20} and \ref{fig:upKakX0100V09800Vf04828a41} show the $K \bar{K}$ collision with velocities higher than the $V_C$. For such velocities, the interaction is nearly elastic. In Fig.~\ref{fig:upKakX0100V09800Vf09312a20}, moving kinks bounce back after the collision, while their final velocities have decreased because of the reduction in their kinetic energy due to the excitation of the internal mode and/or energy radiation.

The most interesting difference between the modified $\tilde{\varphi}^{(6)}$ and the standard $\varphi^6$ models is the possibility of annihilating a light  $K \bar{K}$ pair in the up sector and creation of $K \bar{K}$  pair(s) with a higher rest mass in the down sector. For such an event to occur, the initial energy of the colliding kinks must be high enough and the difference in the rest mass of the kinks in the up and down sectors should be small. This happens for very large initial velocity $v_i$ and large enough deformation parameter $a$. Comparison of Figs.~\ref{fig:upKakX0100V09800Vf09312a20} and \ref{fig:upKakX0100V09800Vf04828a41} clearly shows that by taking the initial velocity as $v_i=0.9800$, such an event does not occur for $a=2.0$, because of the large mass difference between kinks in up and down sectors. But with the same initial velocity and taking $a=4.1$, this interesting event happens.  The topological charge of the kink-antikink system is zero. If the center of mass energy is enough large, the kink and antikink in the up sector may be annihilated and then, it is possible that a set of kink-antikink pair(s) is created in the down sector after collision. This phenomenon can be occurred, if the rest mass of colliding kinks is large as possible. Thus, we expect to observe this phenomenon with large values of the deformation parameter (which leads to a larger value for the rest mass of kinks). Fig.~\ref {fig:upKakX0100V09800Vf04828a41} clearly shows this phenomenon. 
\begin{figure*}[!ht]
\begin{center}
  \centering
    \subfigure[$a=3, v_i=0.300, v_f=0.000$ ]{\includegraphics[width=0.45
 \textwidth]{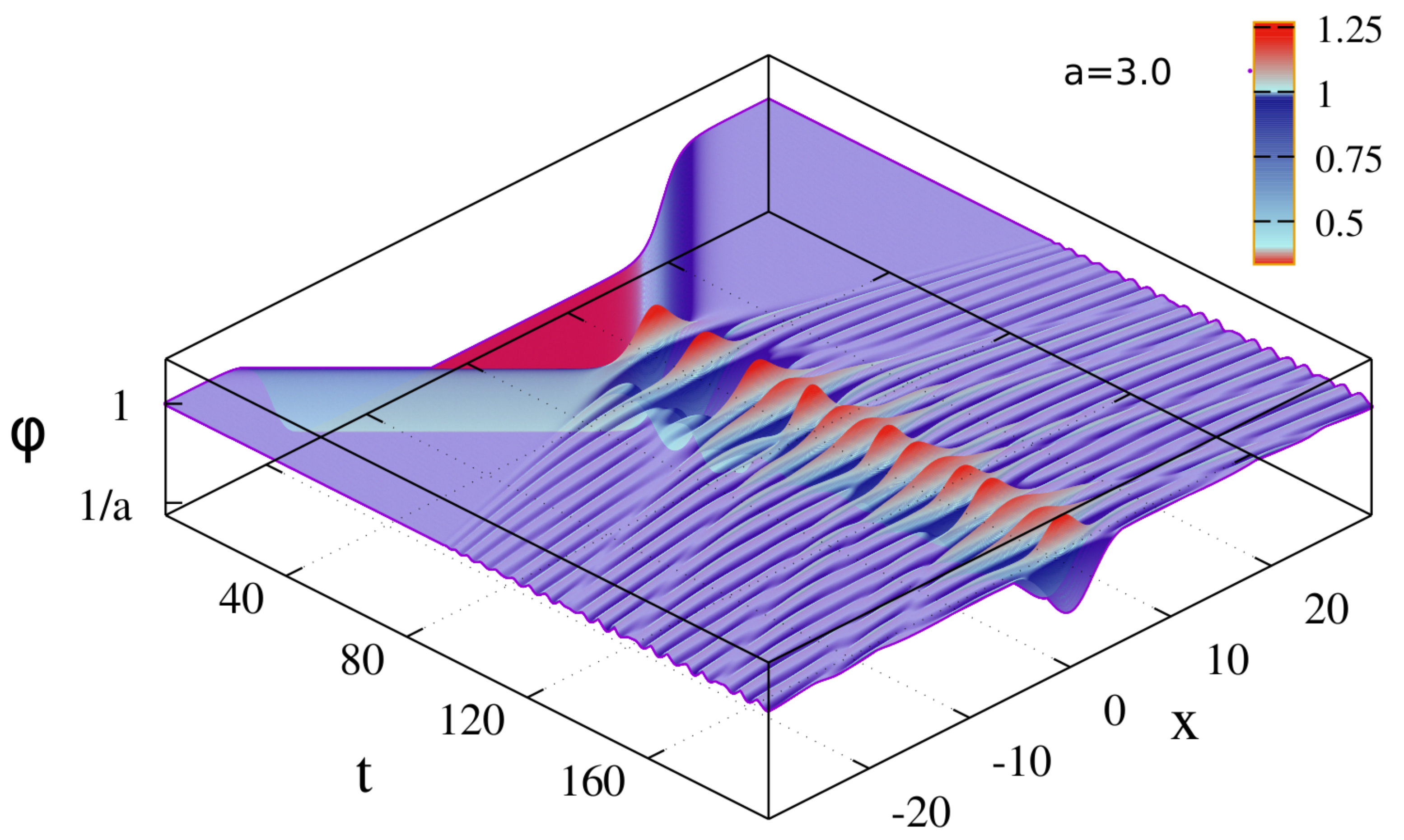}\label{fig:upKakX020V0300a30}}
    \subfigure[$a=3, v_i=0.500, v_f=0.276$ ]{\includegraphics[width=0.45
 \textwidth]{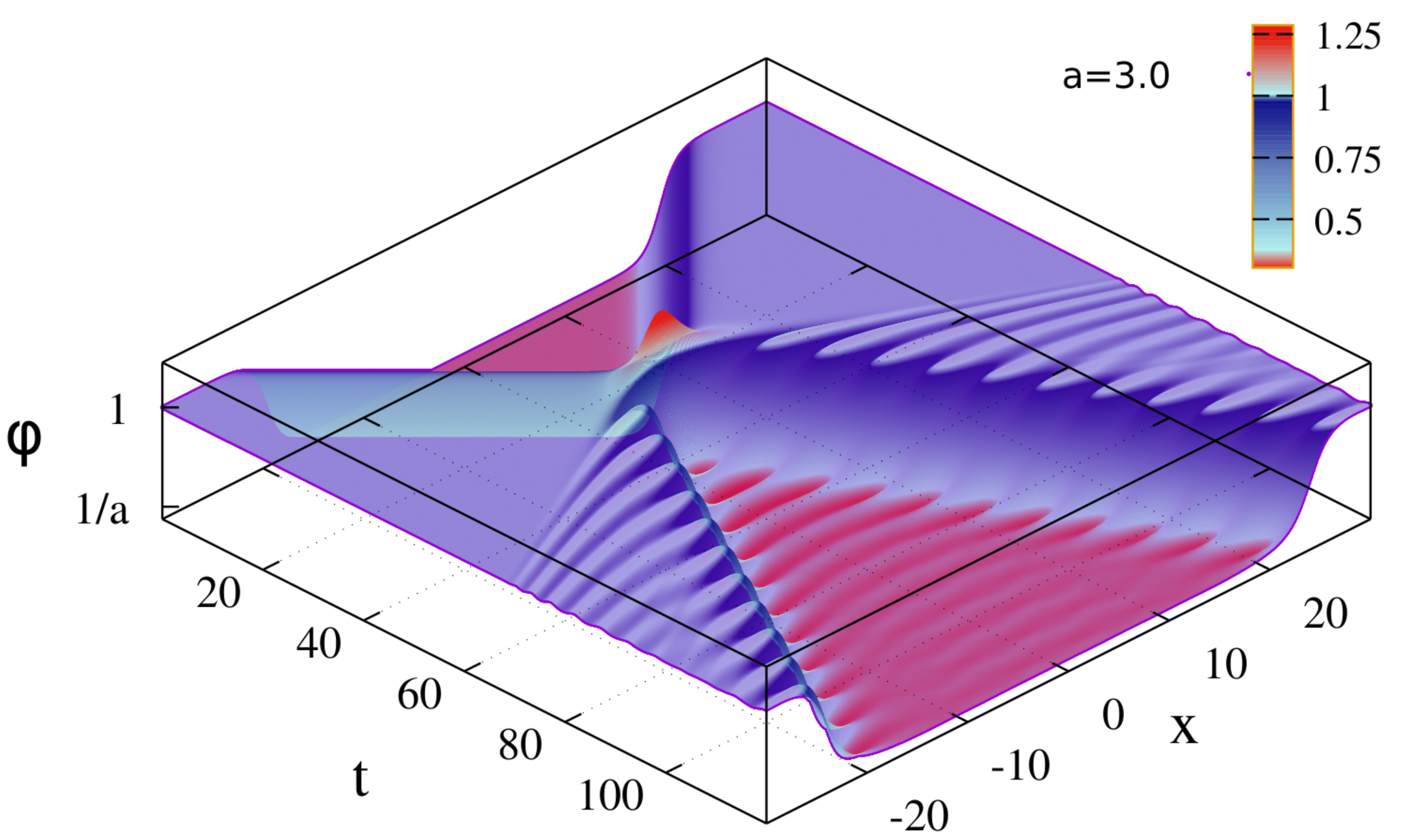}\label{fig:upKakX020V0500Vf0276a30}}
\\
  \caption{Antikink-kink collision in up sector, (a) bion formation for initial velocity less than critical velocity $v_i=0.300 < v_{C}$ and (b) $v_i=0.500 > v_{C}$, the critical velocity for $a=3$ is $v_C=0.406$. }
  \label{fig:upakk}
\end{center}
\end{figure*}

\begin{figure*}[!ht]
\begin{center}
  \centering
    \subfigure[$a=3, v_i=0.73, v_f=0.00$ ]{\includegraphics[width=0.45
 \textwidth]{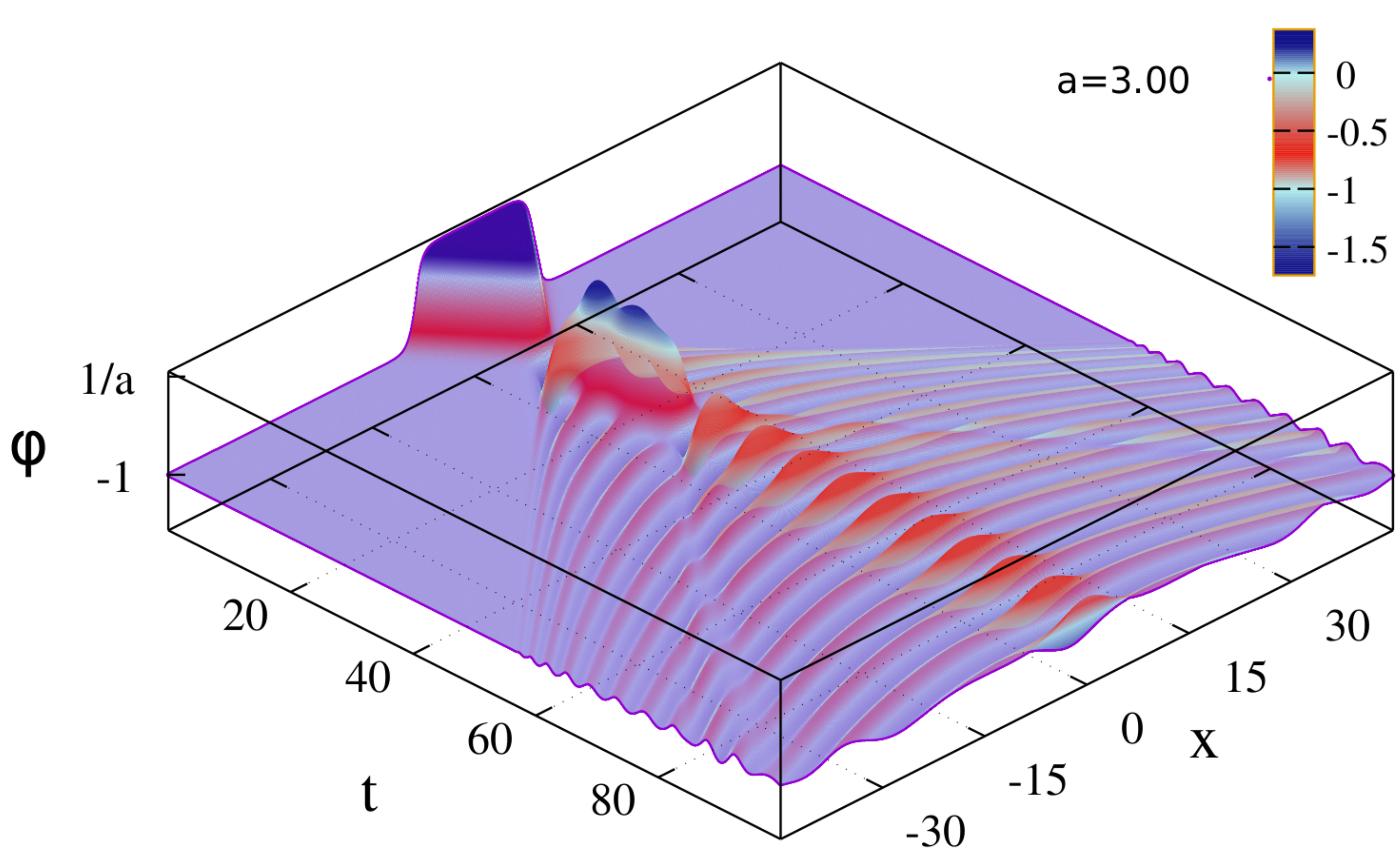}\label{fig:downKakX08V073a300}}
    \subfigure[$a=3, v_i=0.75, v_f=0.13$ ]{\includegraphics[width=0.45
 \textwidth]{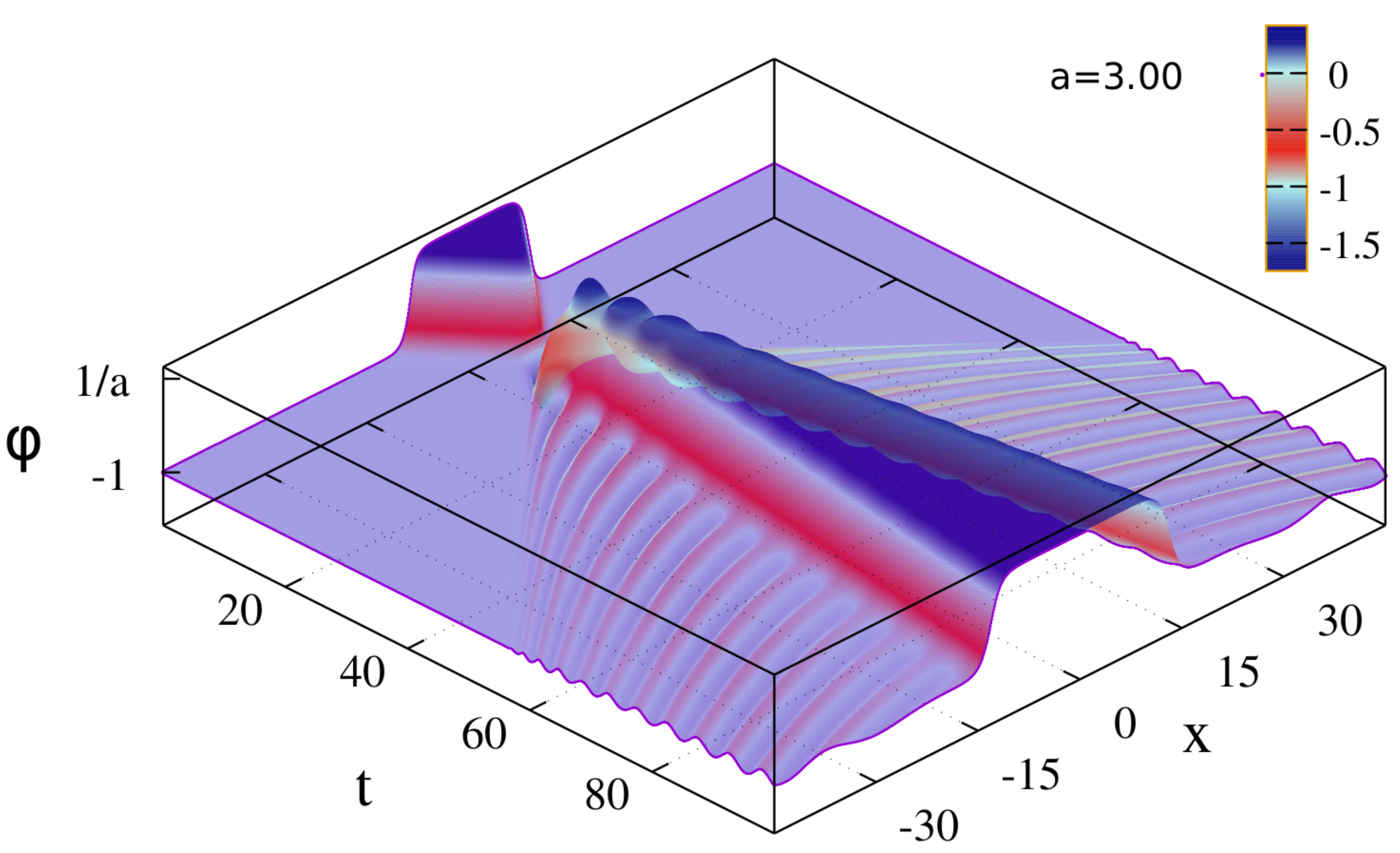}\label{fig:downKakX08V075Vf013a300}}
\\
  \caption{Kink-antikink collision in down sector, (a) bion formation for initial velocity less than critical velocity $v_i=0.7300 < v_{C}$ and (b) $v_i=0.7500 > v_{C}$, the critical velocity for $a=3$ is $v_C=0.7422$. }
  \label{fig:downkak}
\end{center}
\end{figure*}
According to Fig.~\ref{fig:CriticalVelocityUp}  the final state of a $\bar{K}K$ interaction (for any value of the deformation parameter) can be the creation of a bound state ($v_i \le V_C$) or only an almost elastic collision ($v_i > V_C$). The boundary between these two behaviors is determined by the critical velocity $V_C$, which is a function of the deformation parameter $a$. Figs.~\ref{fig:upakk} demonstrate these two different mechanisms for $a=3.0$. By taking the initial velocity $v_i=0.30$, which is lower than the critical velocity $V_C$, the interaction result is the creation of a static bion (see Fig.~\ref{fig:upKakX020V0300a30}). But for initial velocity $v_i=0.5 \ge V_C$ (Fig.~\ref{fig:upKakX020V0500Vf0276a30}), the $\bar{K}K$ interaction ends only in an almost elastic collision. The energy radiation and effects of internal mode interaction are clearly observed in both situations.

The down sector kinks are heavier than the up sector ones for all values of the deformation parameter. Fig.~\ref {fig:CriticalVelocityDown} shows that, like the $K\bar{K}$ collision in the up sector, two different mechanisms occur for similar interaction in the down sector: 1) the establishment of bion bound states (like Fig.~\ref{fig:downKakX08V073a300}) or oscillation in bounce windows for $v_i \le V_C$, or 2) the  nearly elastic collision with $v_i > V_C $ (Fig.~\ref{fig:downKakX08V075Vf013a300}).

\begin{figure*}[t!]
\begin{center}
  \centering
      \subfigure[$a=1.80$ ]{\includegraphics[width=0.32
 \textwidth]{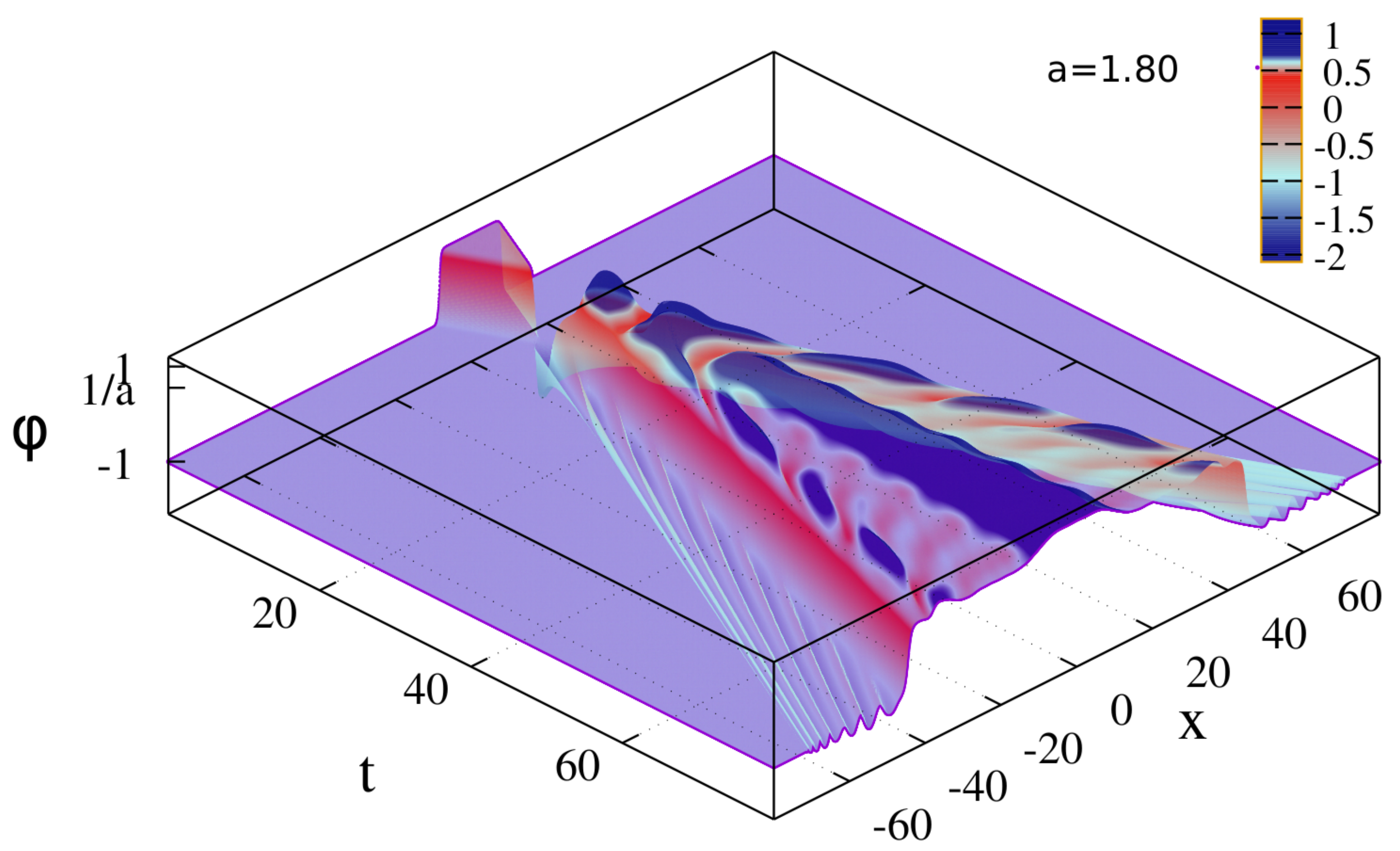}\label{fig:downKakX08V095a180}}
    \subfigure[$a=1.90$ ]{\includegraphics[width=0.32
 \textwidth]{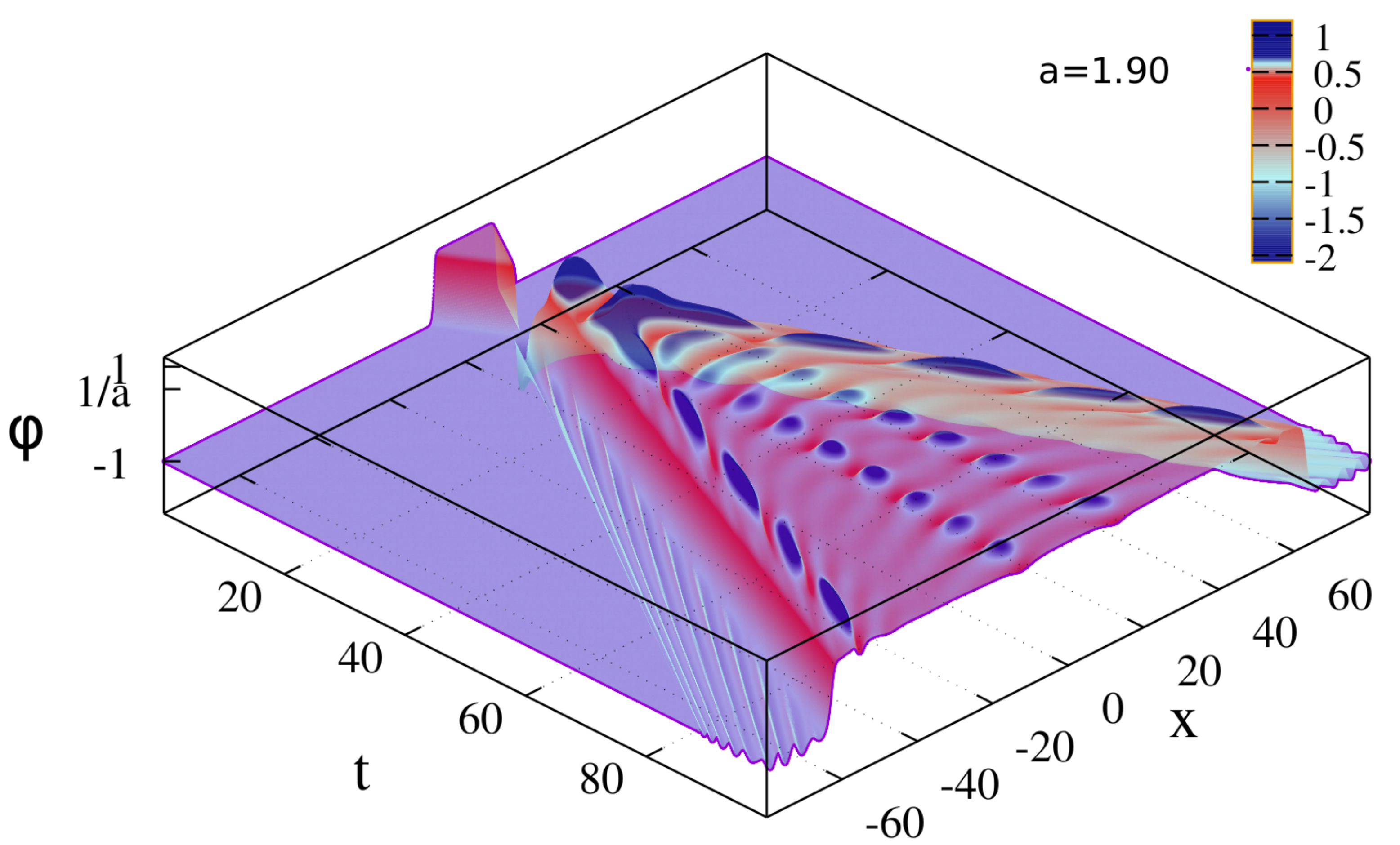}\label{fig:downKakX08V095a190}}
    \subfigure[$a=2.15$ ]{\includegraphics[width=0.32
 \textwidth]{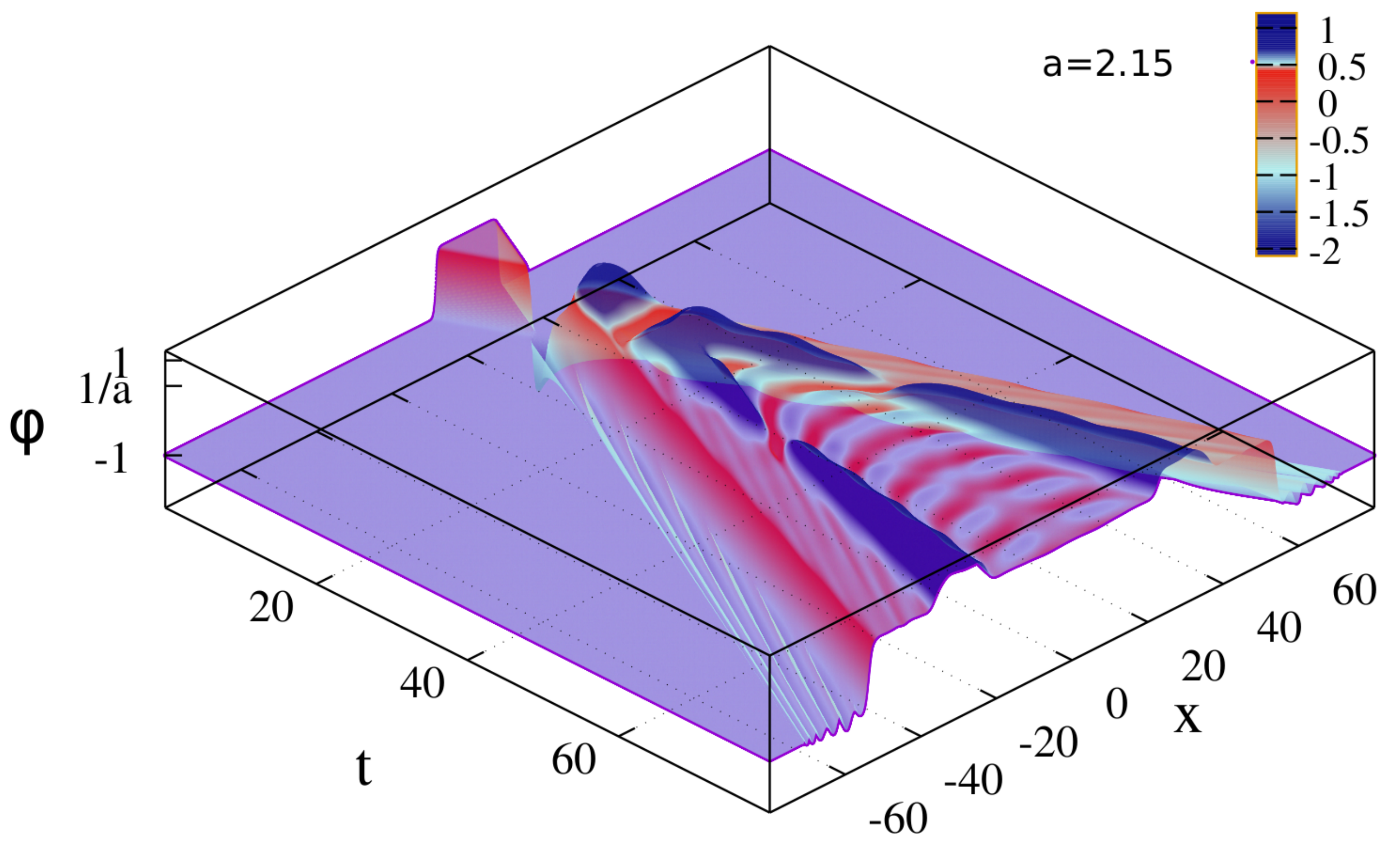}\label{fig:downKakX08V095a215}}
\\
    \subfigure[$a=2.30$ ]{\includegraphics[width=0.32
 \textwidth]{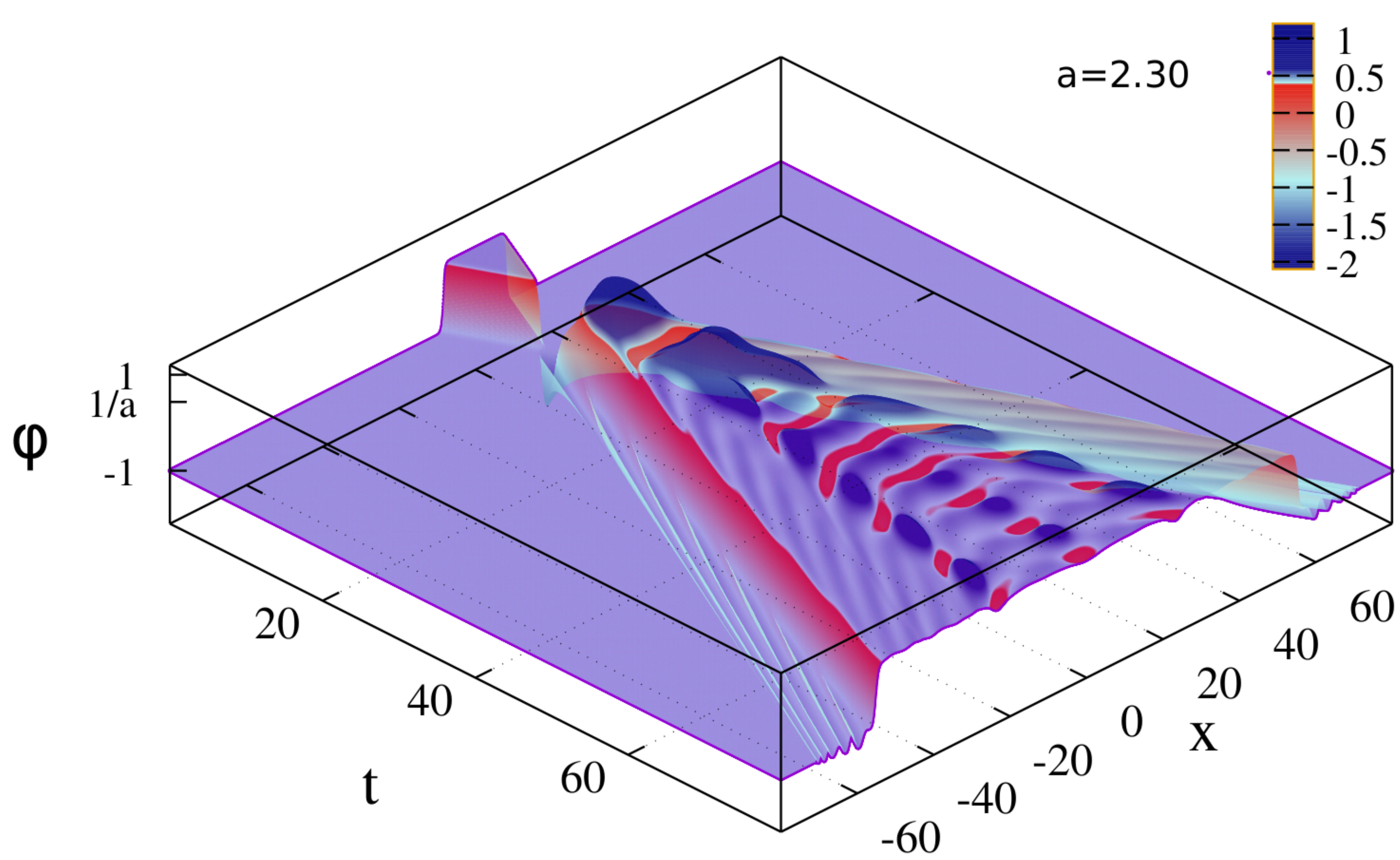}\label{fig:downKakX08V095a230}}
    \subfigure[$a=2.40$ ]{\includegraphics[width=0.32
 \textwidth]{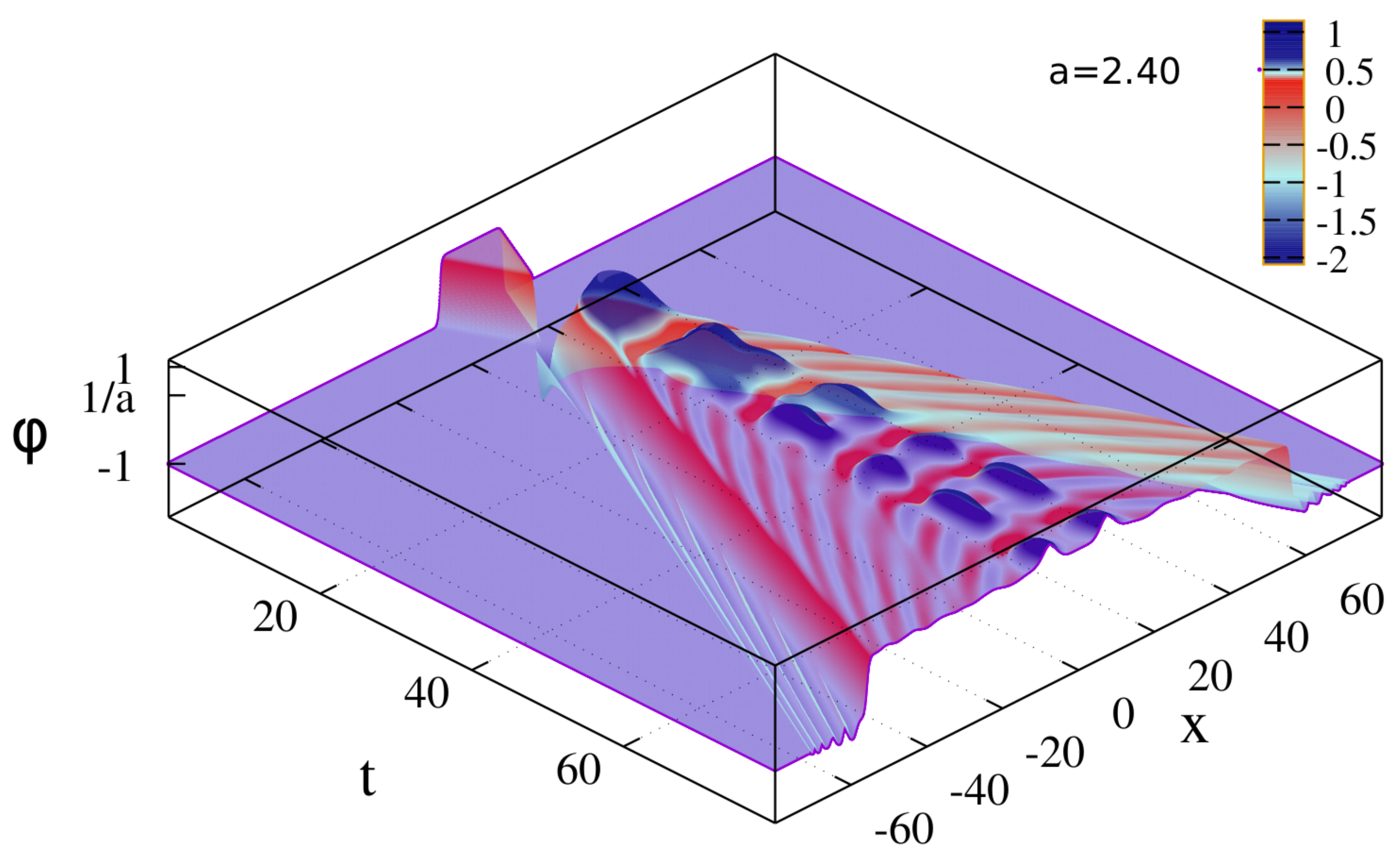}\label{fig:downKakX08V095a240}}
    \subfigure[$a=2.55$ ]{\includegraphics[width=0.32
 \textwidth]{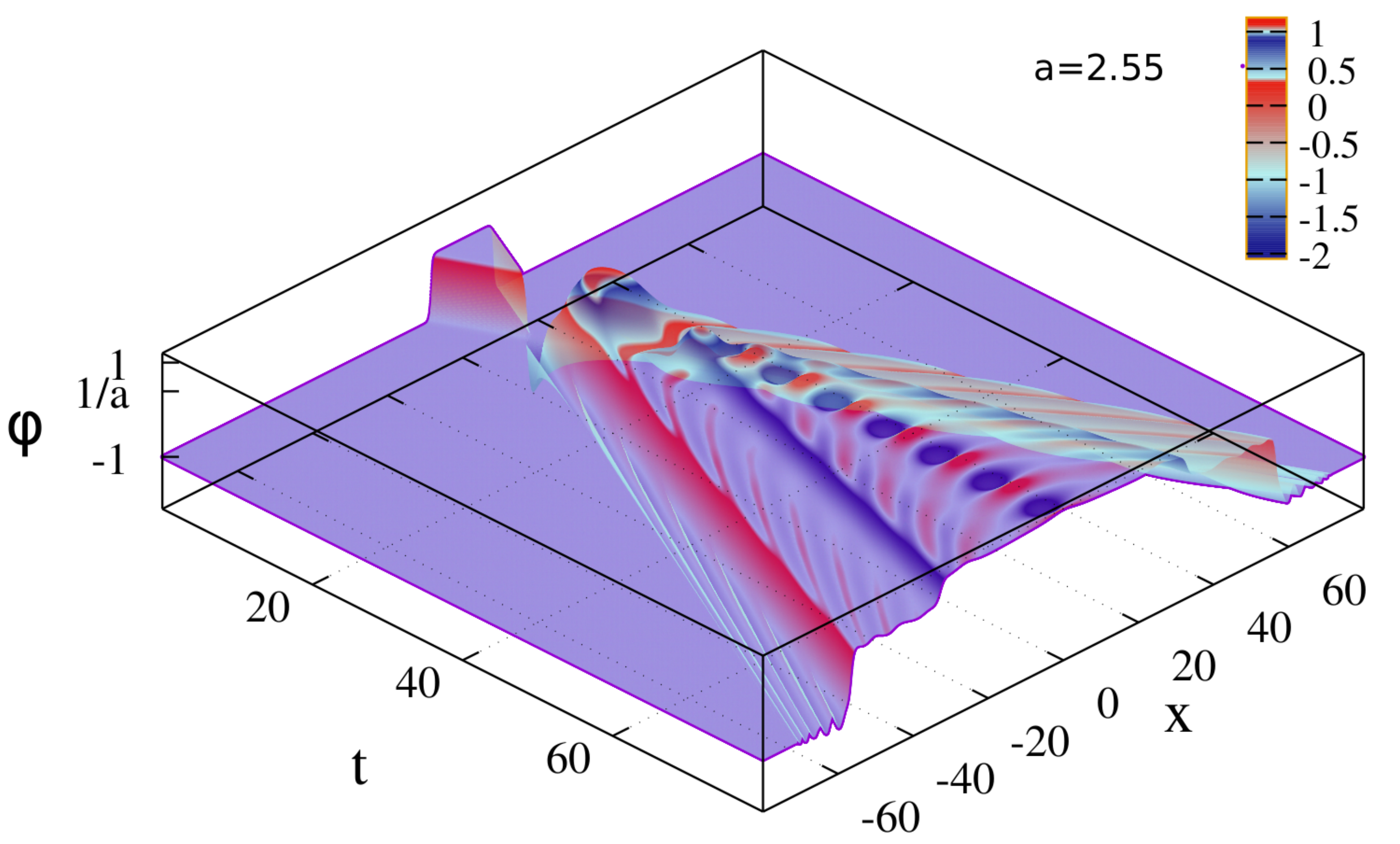}\label{fig:downKakX08V095a255}}
\\
    \subfigure[$a=2.65$ ]{\includegraphics[width=0.32
 \textwidth]{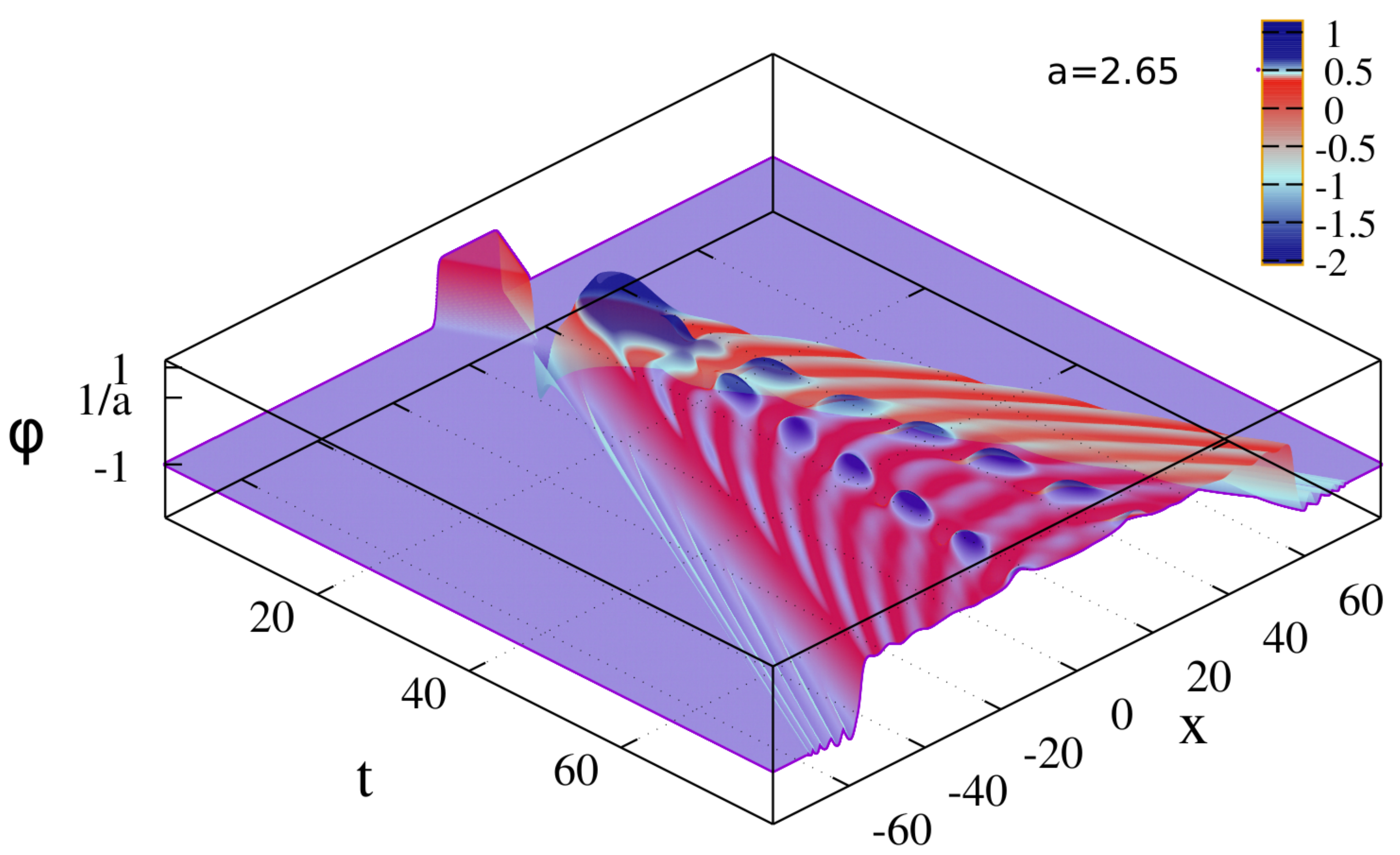}\label{fig:downKakX08V095a265}}
    \subfigure[$a=3.05$ ]{\includegraphics[width=0.32
 \textwidth]{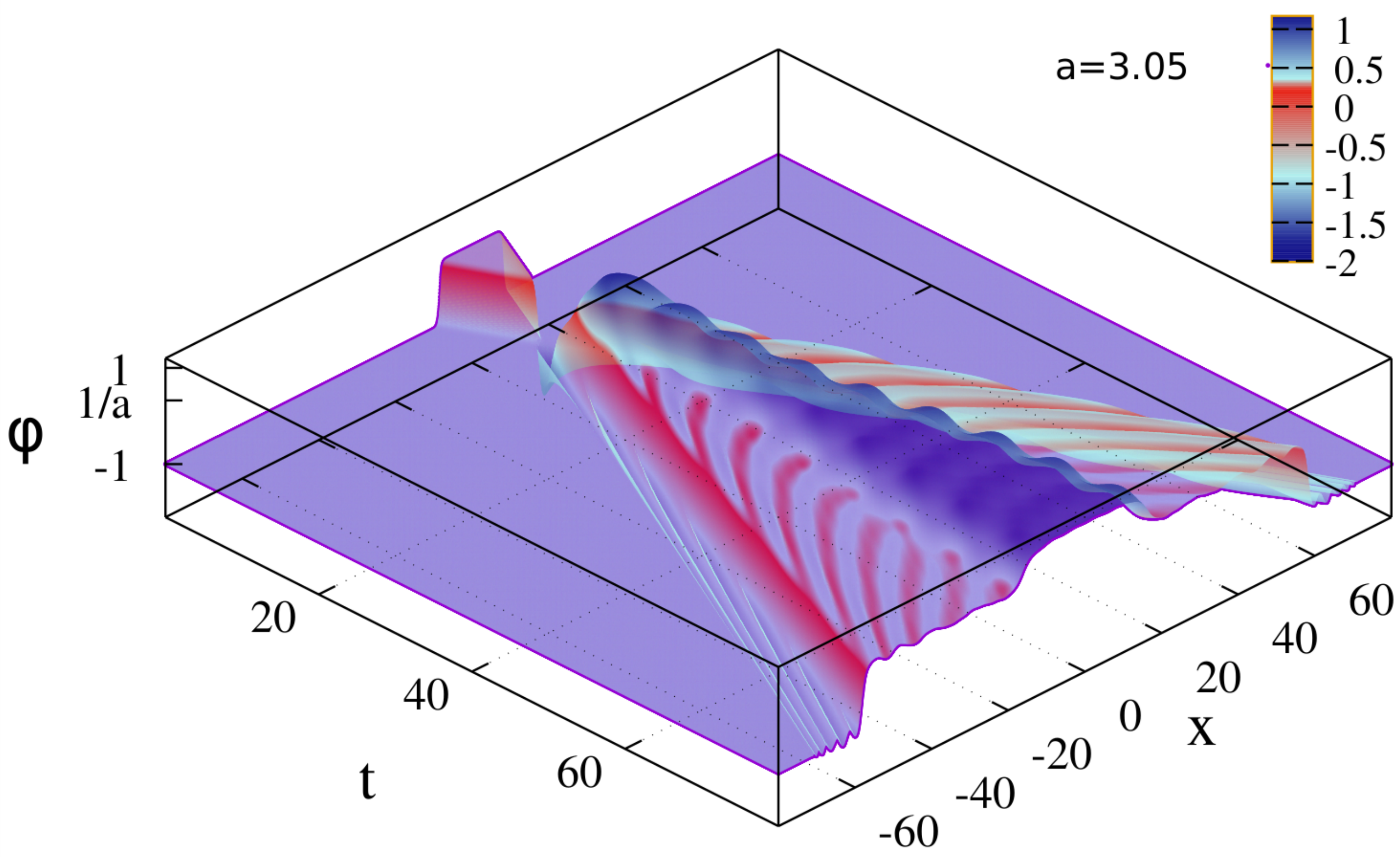}\label{fig:downKakX08V095a305}}
\\
  \caption{The effect of parameter $a$ on obtaining different final products in $K\Bar{K}$ collision, $2X_0=16, v_{i}=0.95 > v_C$ in all figures, (a)  a $K\bar{K}$ pair in new sector and two moving bions, (b) four bions, (c) two $K\bar{K}$ pairs in new sector, (d) three bions, (e) two nearly static bions, (f) a $K\bar{K}$ pair and a bion in new sector, (g) two moving bion and (h)  a $K\bar{K}$ pair in new sector. }
  \label{fig:ChangingSector}
\end{center}
\end{figure*}

For velocities greater than $V_C$, it is possible that the energy transfers to the up sector in the form of bion production or kink-antikink pair(s) creation, due to the larger mass of the kinks in the down sector. Figs.~\ref {fig:ChangingSector} demonstrate various configurations of energy transfer to the up sector for $v_i=0.95 >V_C$ and different values of the deformation parameter $a$ (i.e. different values of the mass difference between kinks of the up and down sectors).

\begin{figure*}[!ht]
\begin{center}
  \centering
    \subfigure[$a=1.5$]{\includegraphics[width=0.45
 \textwidth]{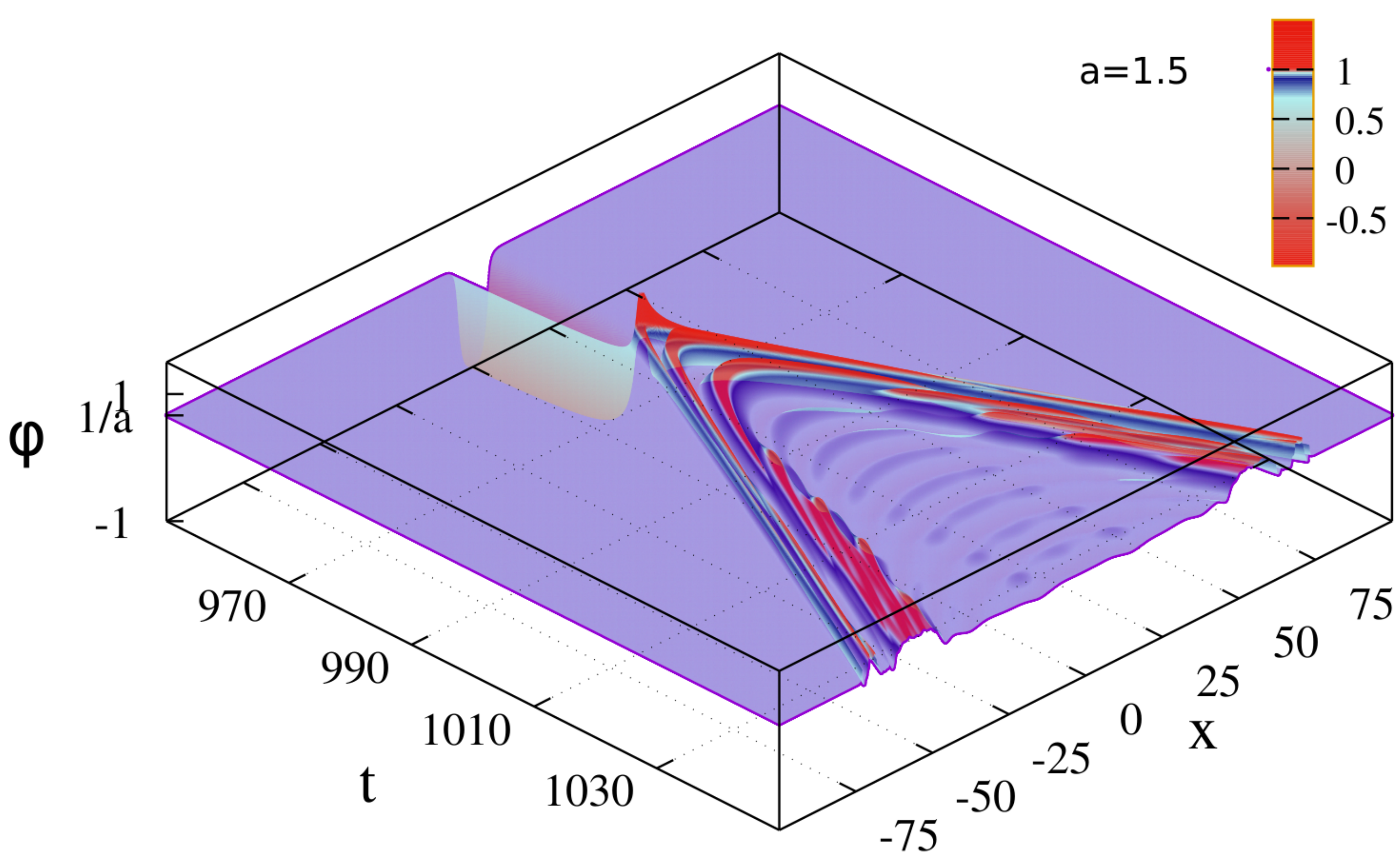}\label{fig:downKakX100V01a15}}
    \subfigure[$a=2.0$ ]{\includegraphics[width=0.45
 \textwidth]{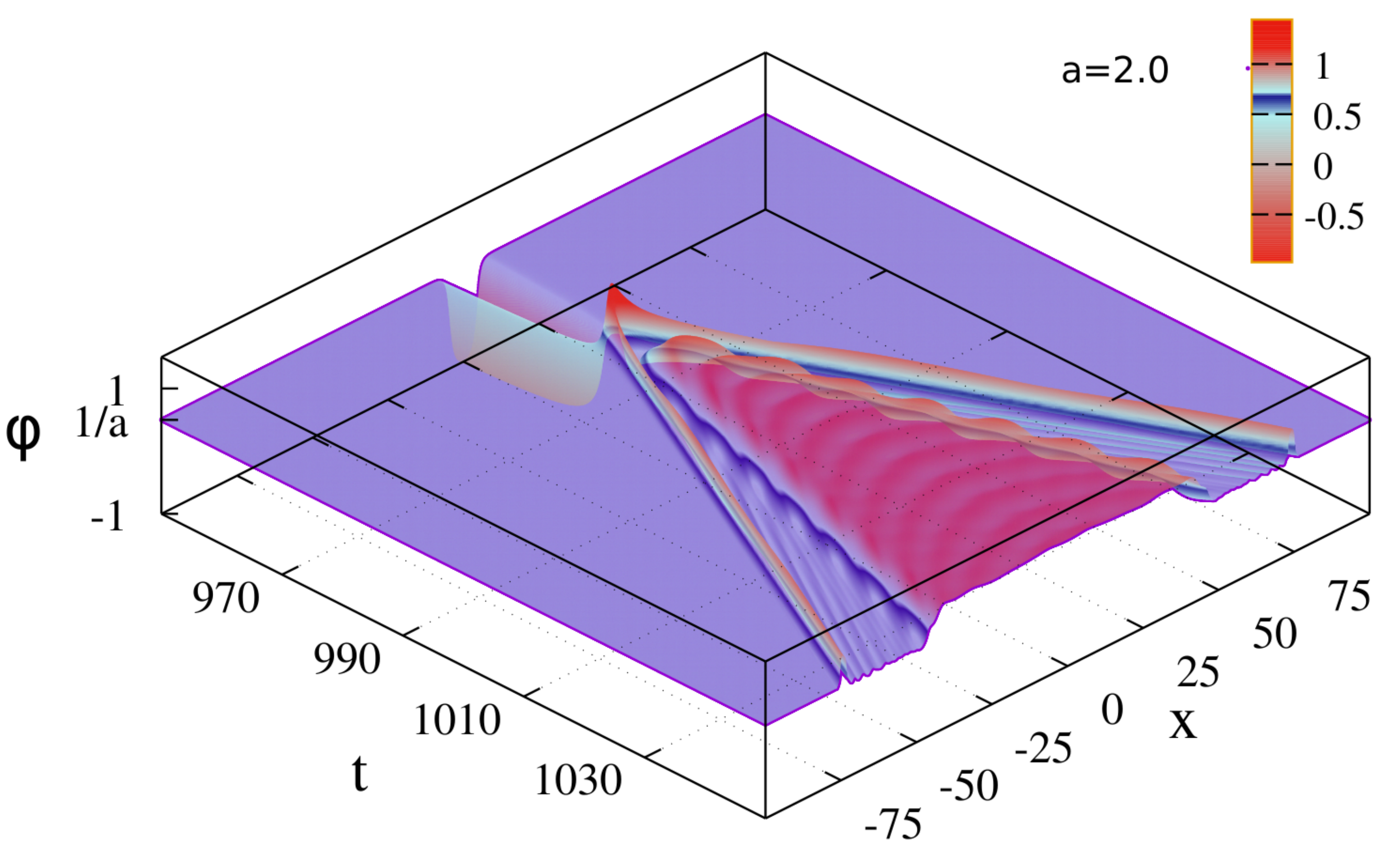}\label{fig:downKakX100V01a20}}
\\
    \subfigure[$a=3.0$]{\includegraphics[width=0.45
 \textwidth]{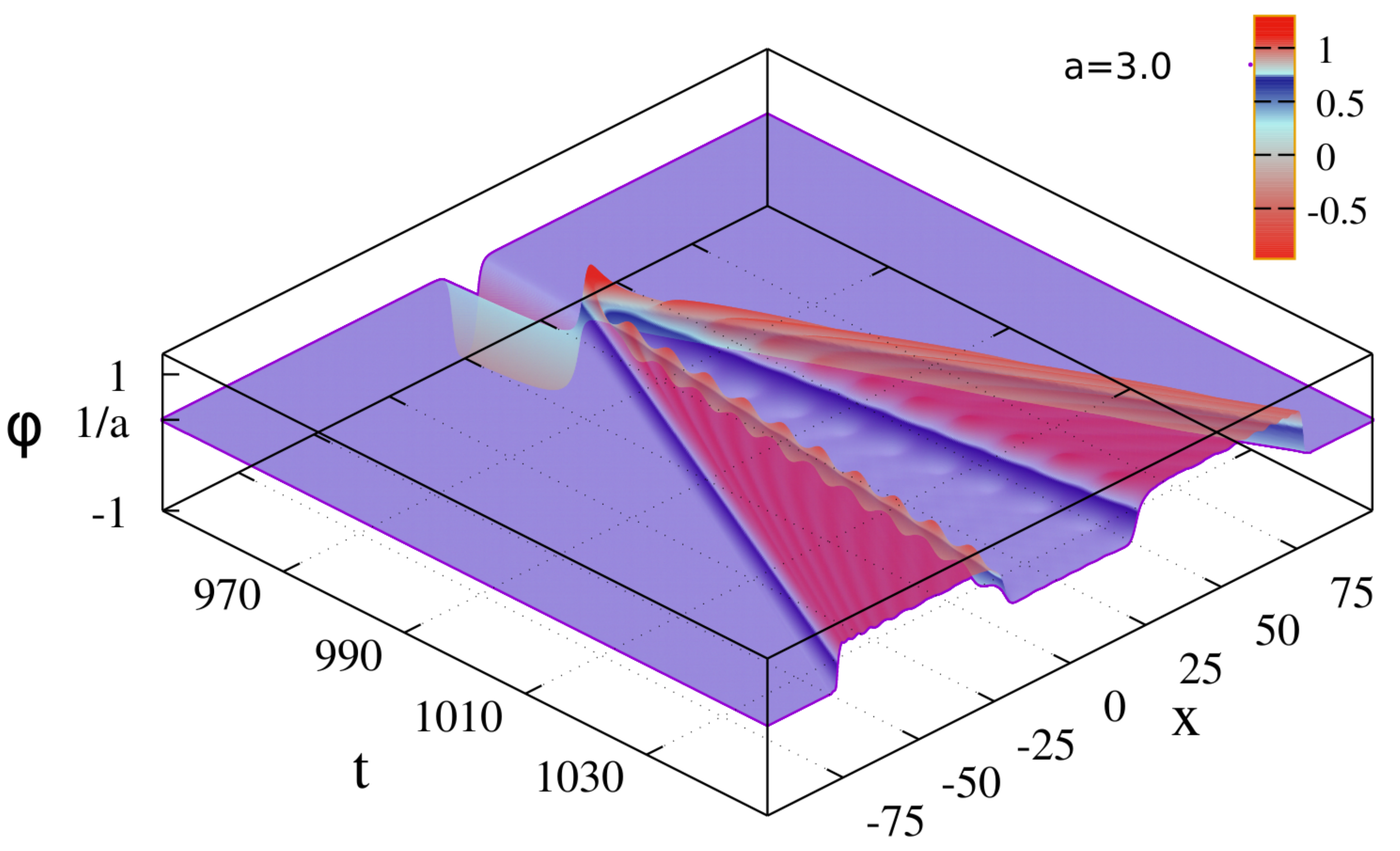}\label{fig:downKakX100V01a30}}
    \subfigure[$a=4.0$ ]{\includegraphics[width=0.45
 \textwidth]{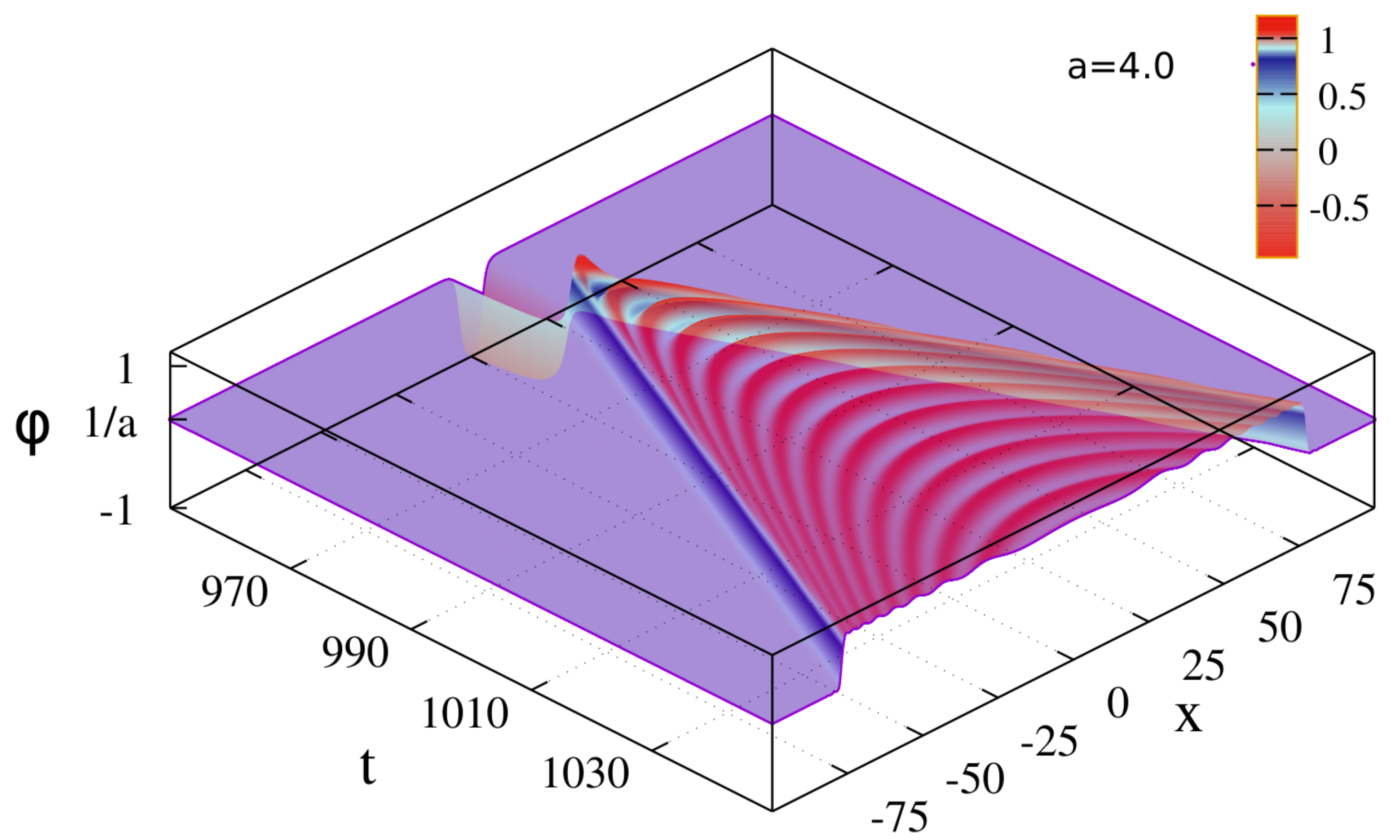}\label{fig:downKakX100V01a40}}
\\
  \caption{Antikink-kink collision in the down sector. In all cases $v_i=0.1$ and $2X0=200$. (a) four pairs, (b) three pairs, (c) two pairs, and (d) a pair kink-antikink formation in the up sector.}
  \label{fig:downakk}
\end{center}
\end{figure*}

The $\bar{K}K$ collision in the down sector has a straightforward mechanism. The colliding kinks are destroyed in the down sector and a number(s) of kink and antikink pair(s) and/or bion(s) are created in the up sector. Figs.~\ref{fig:downakk} demonstrate the different sets of objects created in the $\bar{K}K$ collision with an initial velocity of $v_i=0.1$ and different values of the deformation parameter $a$. As the deformation parameter increases, the mass of the kink in the down sector decreases while the mass of the kink in the up sector increases. As a result, the possibility of creating more $K\bar{K}$ pairs in the up sector reduces.


\begin{figure*}[!ht]
\begin{center}
  \centering
    \subfigure[ ]{\includegraphics[width=0.32
 \textwidth, height=0.185 \textheight]{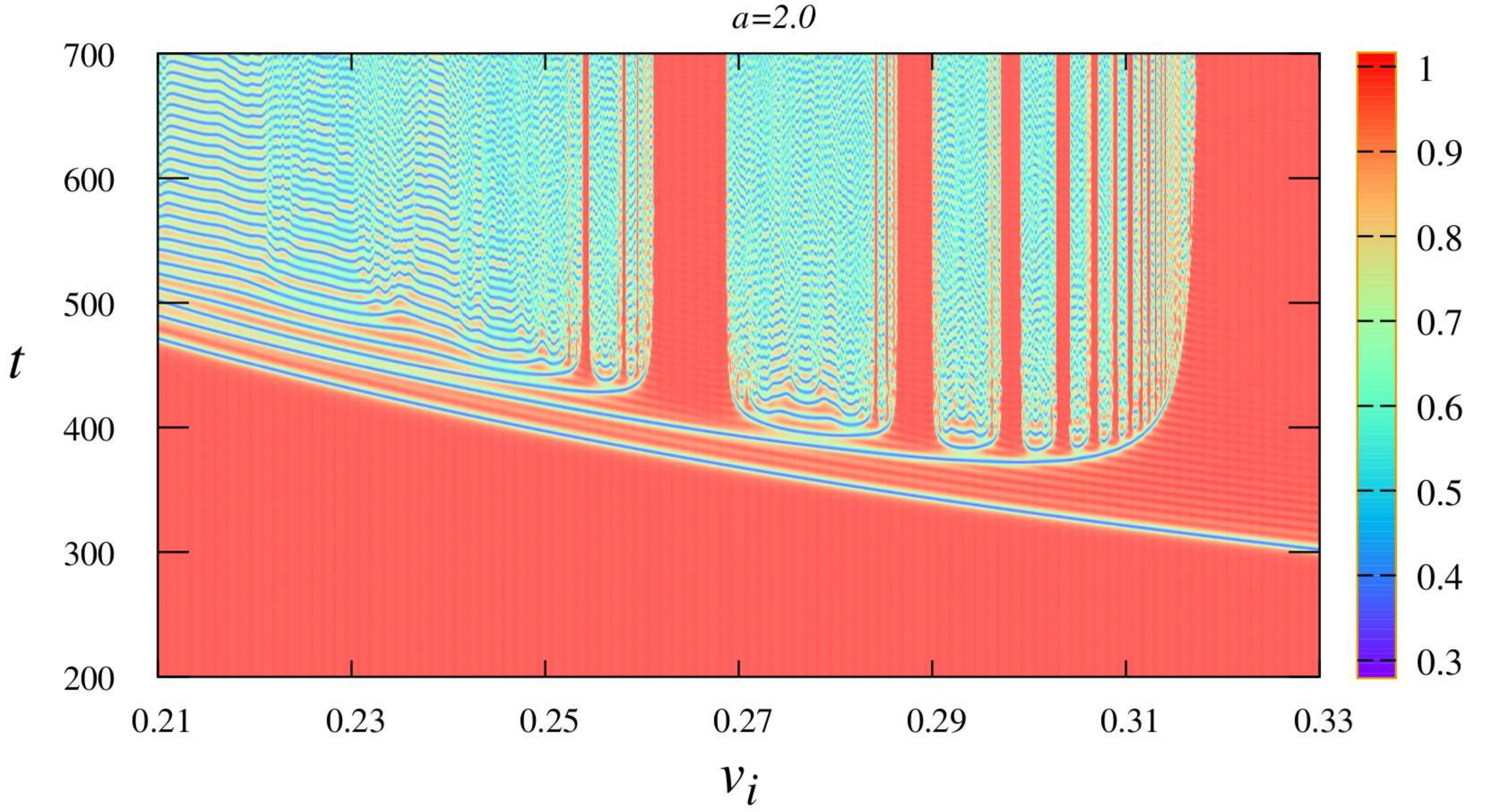}\label{fig:KAKa20Up}}
    \subfigure[ ]{\includegraphics[width=0.32
 \textwidth, height=0.175 \textheight]{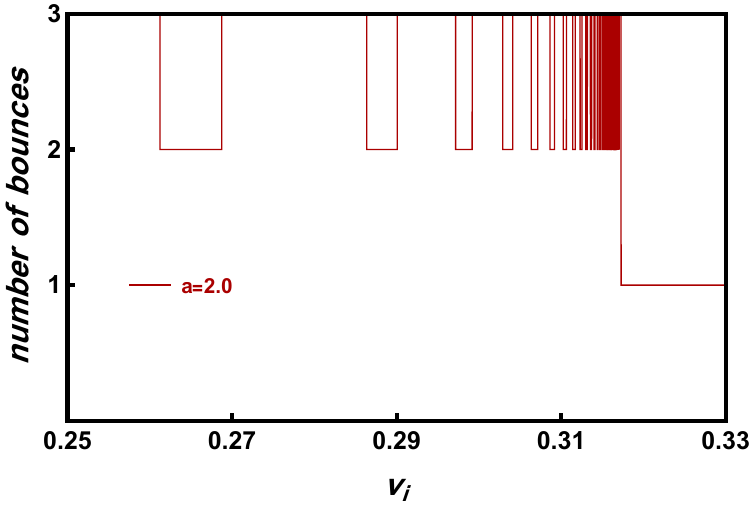}\label{fig:NbUpKAKa20}}
    \subfigure[ ]{\includegraphics[width=0.32
 \textwidth, height=0.170 \textheight]{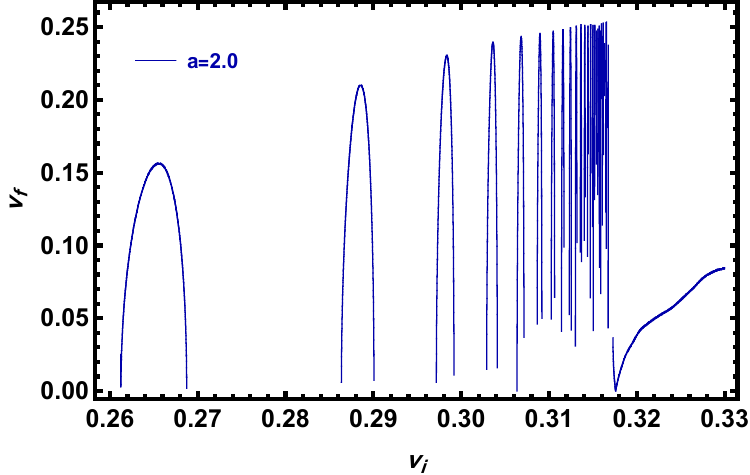}\label{fig:VfUpKAKa20}}
\\
    \subfigure[ ]{\includegraphics[width=0.32
 \textwidth, height=0.185 \textheight]{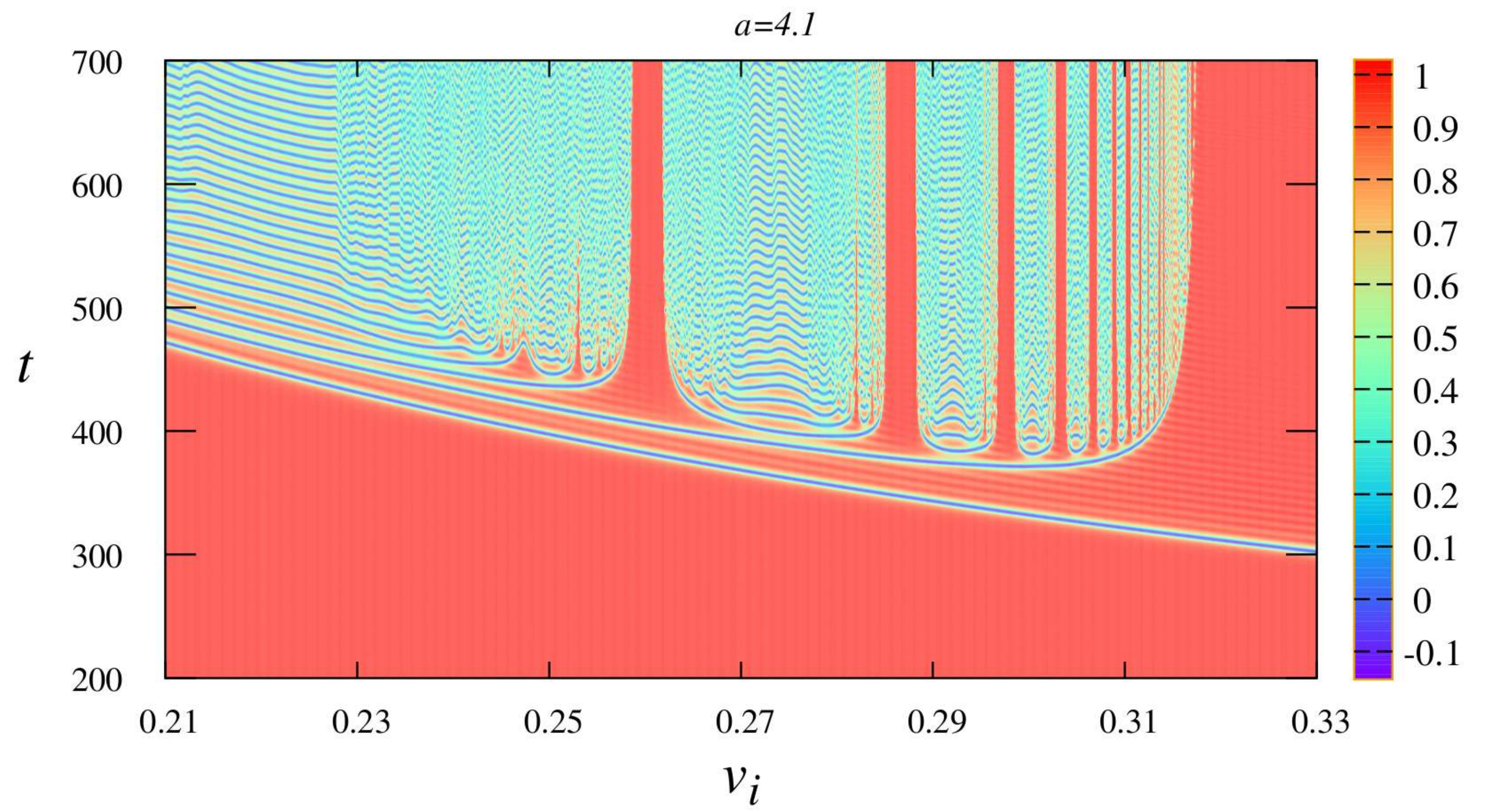}\label{fig:KAKa41Up.pdf}}
    \subfigure[ ]{\includegraphics[width=0.32
 \textwidth, height=0.175 \textheight]{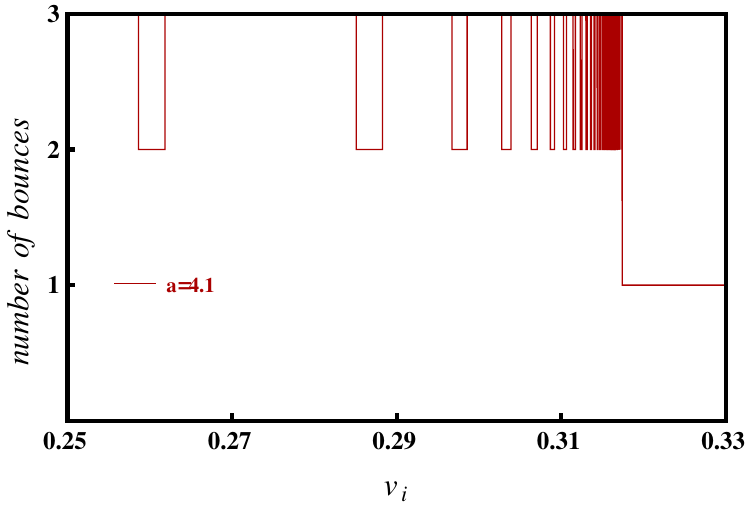}\label{fig:NbUpKAKa41.pdf}}
   \subfigure[ ]{\includegraphics[width=0.32
 \textwidth, height=0.170 \textheight]{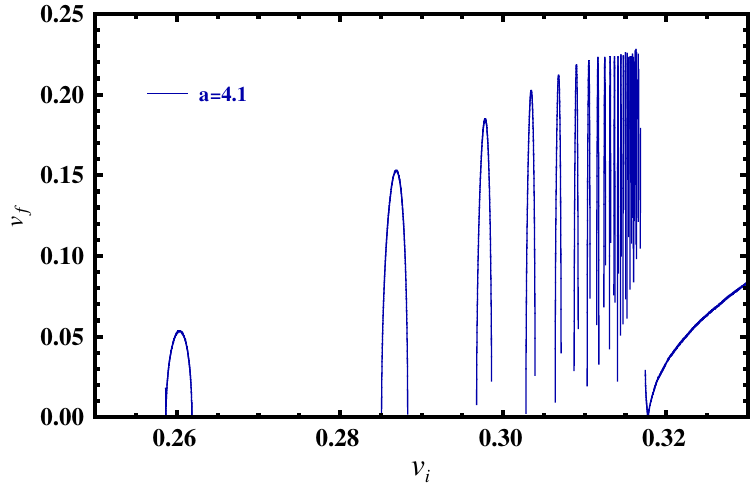}\label{fig:VfUpKAKa41.pdf}}
\\
  \caption{Kink-antikink collision in the up sector, (a), (b), and (c) correspond to simulations with parameter $a=2$ while (d), (e), and (f) correspond to $a=4.1$. In (a) and (d), the plots illustrate the evolution of the scalar field $\tilde{\phi}^{(6)}(\frac1a,1)$ at the center of mass over time and initial velocity. (b) and (e) depict the number of bounces as a function of initial velocity. Finally,  (c) and (f) present the final velocity for cases with one and two bounces as a function of initial velocity, $X_0=100$}
  \label{fig:KaKUpCM}
\end{center}
\end{figure*}

 The complex nature of nonlinear systems makes it difficult to distinguish ordered behaviors from chaotic effects. Although finding chaotic behaviors in a nonlinear system can guide us to find similar behaviors in another nonlinear models, we cannot generalize to all similar systems. The critical velocity gives a suitable criterion for the emergence of chaotic behavior in a certain region of the system initial conditions. But what is the effect of deformation on such behaviors? As can be seen from Fig.~\ref {fig:CriticalVelocityUp}, for the ${K \bar{K}}$ collision in the up sector of deformed model $\Tilde{\varphi}^{(6)}$, we may find same critical velocity for two different values of the deformation parameter $"a"$. For example, by taking $a=2.0$ and $a=4.1$, the critical velocity becomes $v_c=0.318$. Fig.~\ref {fig:KAKa20Up} (\ref{fig:KAKa41Up.pdf}) shows time evolution of the kink location for different values of initial velocity with deformation parameter $a=2.0$ ($a=4.1$). Several bounce windows that occur for initial speeds lower than the critical velocity are well recognized in these figures. One can observe the chaotic nature of the final state in the $K\bar{K}$ interaction in figure set \ref {fig:KaKUpCM}. The kink rest mass in the up sector (according to (\ref {eq:modifiedphi6mass})) and the internal mode eigen frequency (see Fig.~\ref {fig:modifiedphi6modes}) increase as the deformation parameter $a$ increases. Bounce windows for larger values of deformation parameter (here, $a=4.1$) occur at lower velocities (Fig.~\ref {fig:NbUpKAKa20}) in comparison with similar bounce windows for $a=2.0$ (Fig.~\ref {fig:NbUpKAKa41.pdf}). On the other hand, the escape velocity of colliding kinks with initial velocity within a bounce window for $a=2.0$ (Fig.~\ref {fig:VfUpKAKa20} ) is greater than the kink escape velocities for $a=4.1$ (Fig.~\ref {fig:VfUpKAKa41.pdf}), because of larger rest mass of kinks in deformed models with greater deformation parameter. Please note that we have compared two different values of deformation parameters with same critical velocity. 

\begin{figure*}[!ht]
\begin{center}
  \centering
    \subfigure[ ]{\includegraphics[width=0.32
 \textwidth, height=0.185 \textheight]{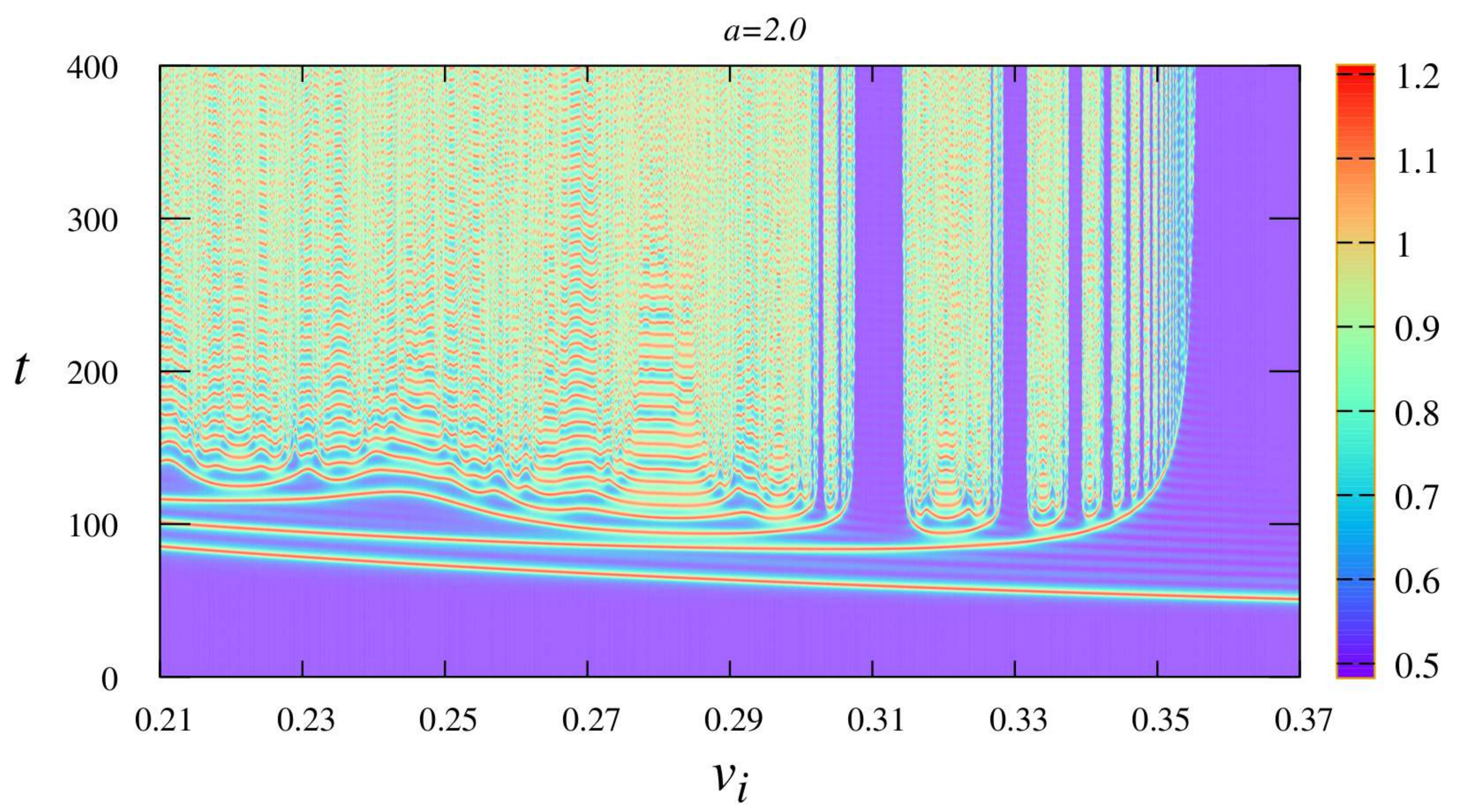}\label{fig:AKKa20Up}}
    \subfigure[ ]{\includegraphics[width=0.32
 \textwidth, height=0.175 \textheight]{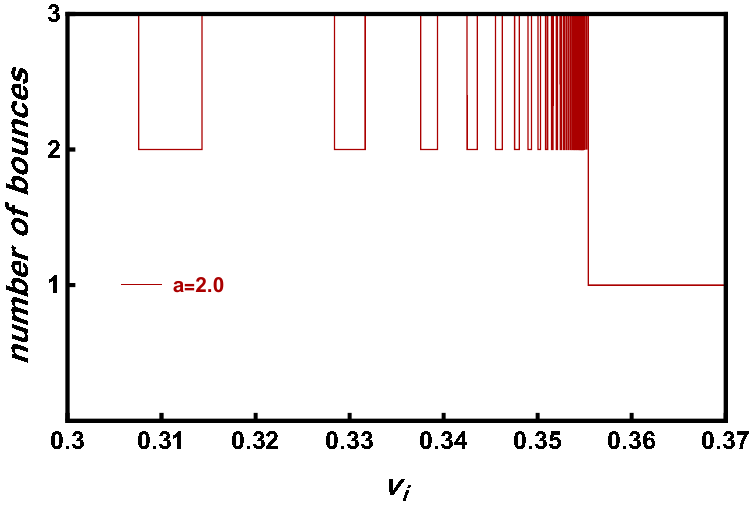}\label{fig:NbUpAKKa20}}
    \subfigure[ ]{\includegraphics[width=0.32
 \textwidth, height=0.170 \textheight]{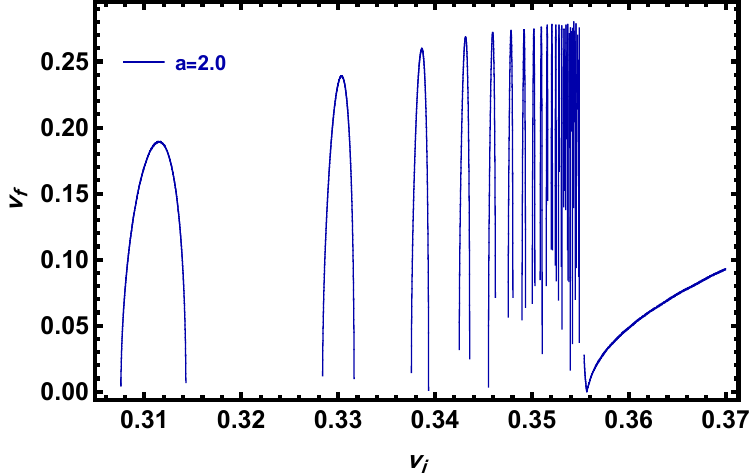}\label{fig:VfUpAKKa20}}
\\
    \subfigure[ ]{\includegraphics[width=0.32
 \textwidth, height=0.185 \textheight]{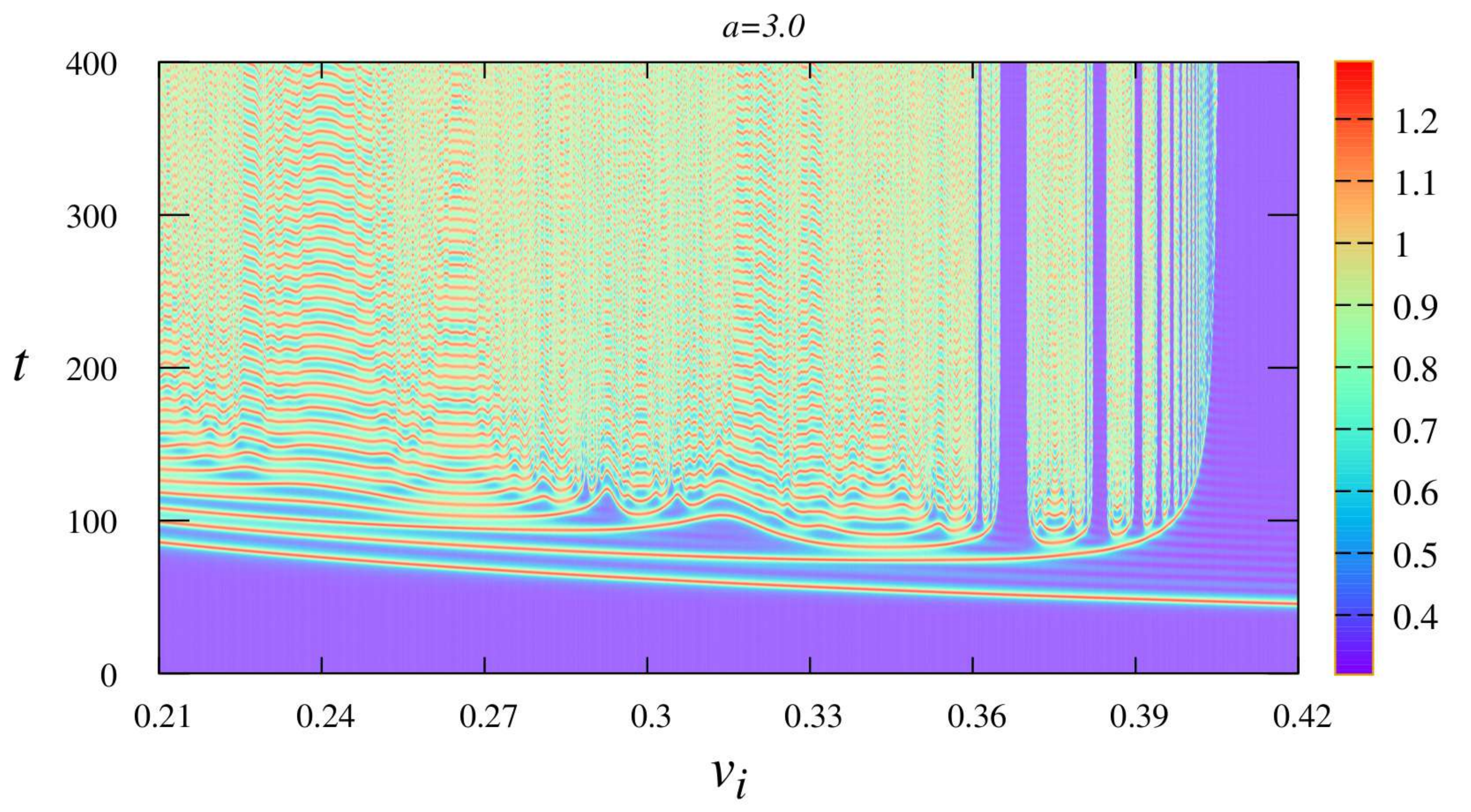}\label{fig:KAKa30Up}}
    \subfigure[ ]{\includegraphics[width=0.32
 \textwidth, height=0.175 \textheight]{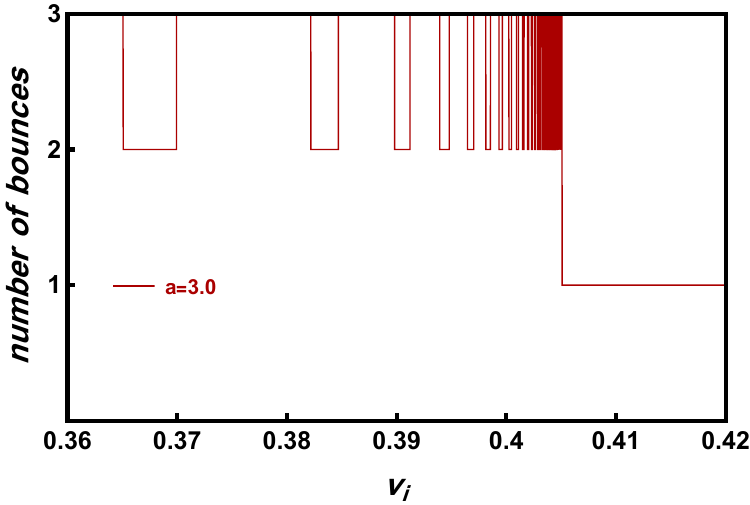}\label{fig:NbUpAKKa30}}
    \subfigure[ ]{\includegraphics[width=0.32
 \textwidth, height=0.170 \textheight]{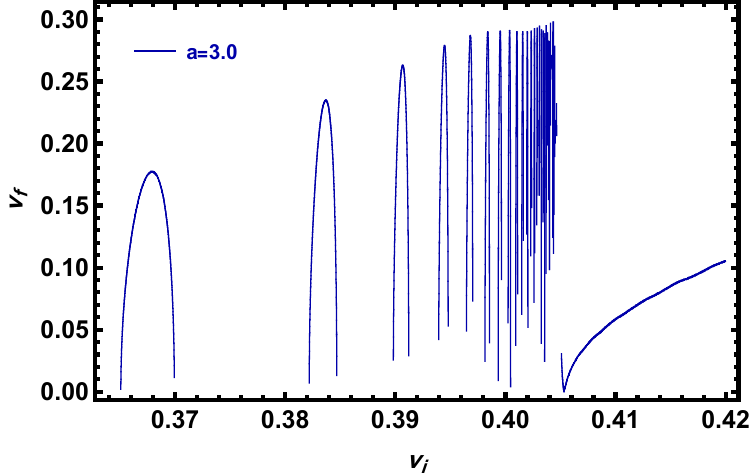}\label{fig:VfUpAKKa30}}
\\
    \subfigure[ ]{\includegraphics[width=0.32
 \textwidth, height=0.185 \textheight]{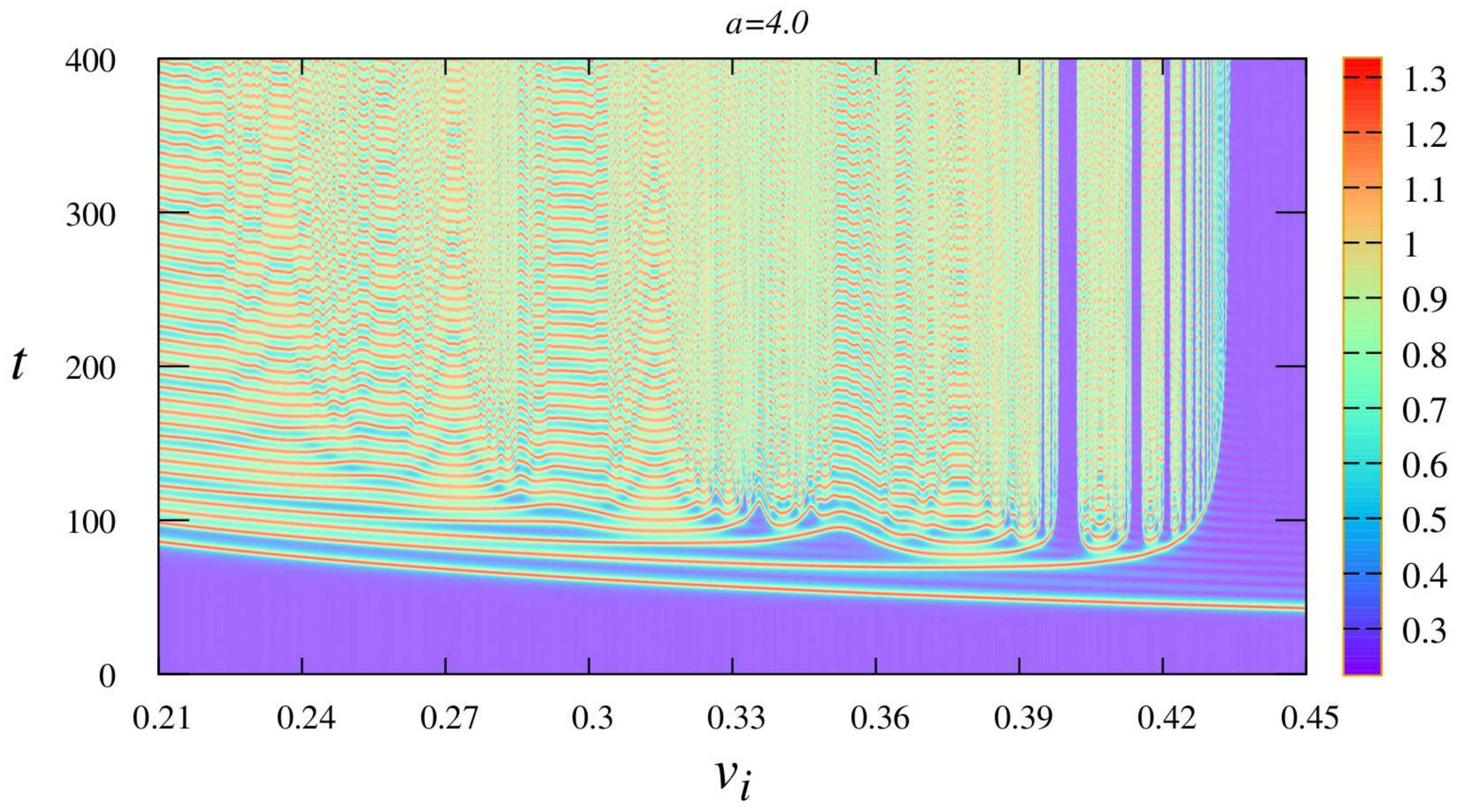}\label{fig:KAKa40Up}}
    \subfigure[ ]{\includegraphics[width=0.32
 \textwidth, height=0.175 \textheight]{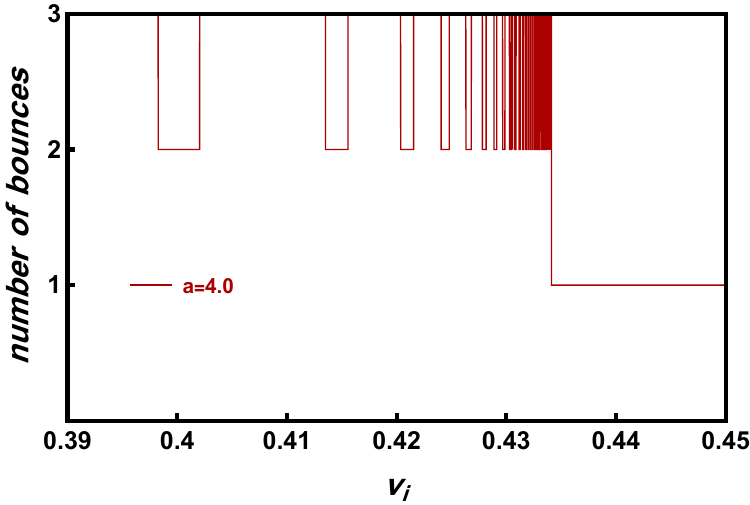}\label{fig:NbUpAKKa40}}
    \subfigure[ ]{\includegraphics[width=0.32
 \textwidth, height=0.170 \textheight]{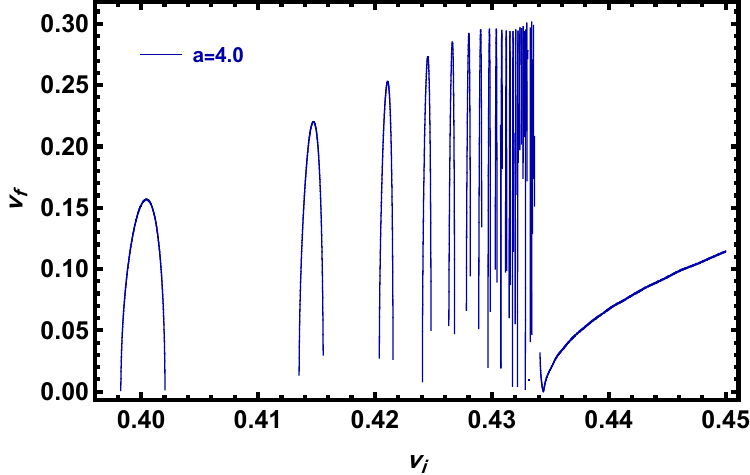}\label{fig:VfUpAKKa40}}
\\
  \caption{antikink-kink collision in the up sector, (a), (b), and (c) correspond to simulations with parameter $a=2$ while (d), (e), and (f) correspond to $a=3$ and (g), (h), and (i) correspond to $a=4$. In (a), (d) and (g), the plots illustrate the evolution of the scalar field $\tilde{\phi}^{(6)}(\frac1a,1)$ at the center of mass over time and initial velocity. (b), (e) and (h) depict the number of bounces as a function of initial velocity. Finally,  (c), (f) and (i) present the final velocity for cases with one and two bounces as a function of initial velocity, $X_0=100$}
  \label{fig:aKKUpCM}
\end{center}
\end{figure*}

The $\bar{K}K$ collision occurs in a different mechanism from the $K\bar{K}$ interaction. Even if the creation of bound states structure in the $\bar{K}K$ collision does not need the existence of an internal mode. This is due to the formation of the potential well through Eq.~(\ref {eq:stabilitypotentials}) in the field mode equation Eq.~(\ref {eq:schrodingerlike} )\cite {dorey}. By inserting field solution Eq.~(\ref {eq:modifiedphi6kinks}) in pseudo-potential Eq.~(\ref {eq:stabilitypotentials}) one can find that the depth of the created potential barrier in the up sector during $\bar{K}K$ interaction  (as well as the rest mass of kink solutions) increases as the deformation parameter increases (see Fig.~\ref {fig:modifiedphi6qmpup}). Thus, the critical velocity $v_c$ increases with an increase of deformation parameter. Figs. \ref {fig:aKKUpCM} demonstrate time evolution of field at the center of mass during $\bar{K}K$ interaction as a function of the initial velocity for different values of the deformation parameter $a$.

The resonance windows can be observed at higher kink initial velocities ($v_i$) with the increase of the deformation parameter $a$, because the critical velocity becomes larger with the increase of the mentioned parameter (compare Figs.~ \ref {fig:KAKa20Up} and \ref {fig:AKKa20Up}). By comparing Figs.~ \ref {fig:NbUpAKKa20} , \ref{fig:NbUpAKKa30}  and \ref {fig:NbUpAKKa40}, one can find another important phenomenon, which is the shift of resonance windows towards larger initial velocities when the deformation parameter increases. This issue appears due to the increase of the kink rest mass.

Figs.~ \ref {fig:VfUpAKKa20}, \ref {fig:VfUpAKKa30} and \ref {fig:VfUpAKKa40} show another fact about the presented deformed model. The kink final velocity after bounce interactions in the resonance windows decreases with the increase of the deformation parameter $a$, although the bounce windows occurred at larger initial velocities with the increase of this parameter. For example, the first two bounce window with $a=2.0$ occurs in the interval $0.308<v_i<0.314$, while kink escape velocity is $v_f\approx 0.19$. For $a=3.0$ ($a=4.0$), the first two bounce window is moved toward the range $0.365<v_i<0.37$ ($0.398<v_i<0.402$) with kink final velocity $v_f \approx 0.18$ ($v_f \approx 0.16$).

\section {Collective analysis}
\label{sec:Collective analysis}
One can consider the center of each kink (antikink) as the position of a pseudo-particle which moving under the influence of collision processes, kink-internal mode interactions and possibly external force \cite{Takyi.PRD.2016, Kevrekidis.book.2019, Manton.PRL.2021,Adam.PRD.2022}. A suitable set of collective coordinates (in 1+1 dimensions) can be constructed by inserting initially well separated single soliton solutions $\varphi_i(x, X_i)$ (with positions $X_i, i=1,…,N$) into the field action density $\mathcal{L}(t, x, \dot{x}, X_i, \dot{X}_i)$ and integrating over spatial coordinate $x$ to find an effective action $L(t, X_i(t), \dot{X}_i(t)$. Trajectory of moving kinks are derived by solving the Euler-Lagrange equations obtained from the effective Lagrange $L(t, X_i(t), \dot{X}_i(t)$. Such process can be analytically calculated for some field theories with simple potentials. Unfortunately, there is no analytical solution for the introduced deformed model. We have used an alternative method to investigate the collective behavior of kink scattering in deformed model. 

In the $K\bar{K}$ ($\bar{K}K$) system the collective coordinate X(t) represents midpoint of the distance between the kink and the antikink. We define the collective position of the antikink (kink) as \cite{Weigel.JPCS.2014}:
\begin{equation}\label{eq:collective position}
\left\langle x \right\rangle (t)=\frac{\int_0^\infty x \mathcal{H}(x,t)}{\int_0^\infty \mathcal{H}(x,t)}
\end{equation}
where the Hamiltonian density of the system is:
\begin{eqnarray}\label{eq:hamiltonian density}
\mathcal{H} &=& \left(\frac{1}{2}\left(\frac{\partial \Tilde{\varphi}^{(6)}}{\partial t}\right)^2+\frac{1}{2}\left(\frac{\partial \Tilde{\varphi}^{(6)}}{\partial x}\right)^2+\tilde{V}(\Tilde{\varphi}^{(6)})\right). 
\end{eqnarray}
with different initial configurations $K\bar{K}$ (or $ \bar{K}K$ ) which has been explained in previous section.

\begin{figure*}[!ht]
\begin{center}
  \centering
    \subfigure[ ]{\includegraphics[width=0.45
 \textwidth, height=0.185 \textheight]{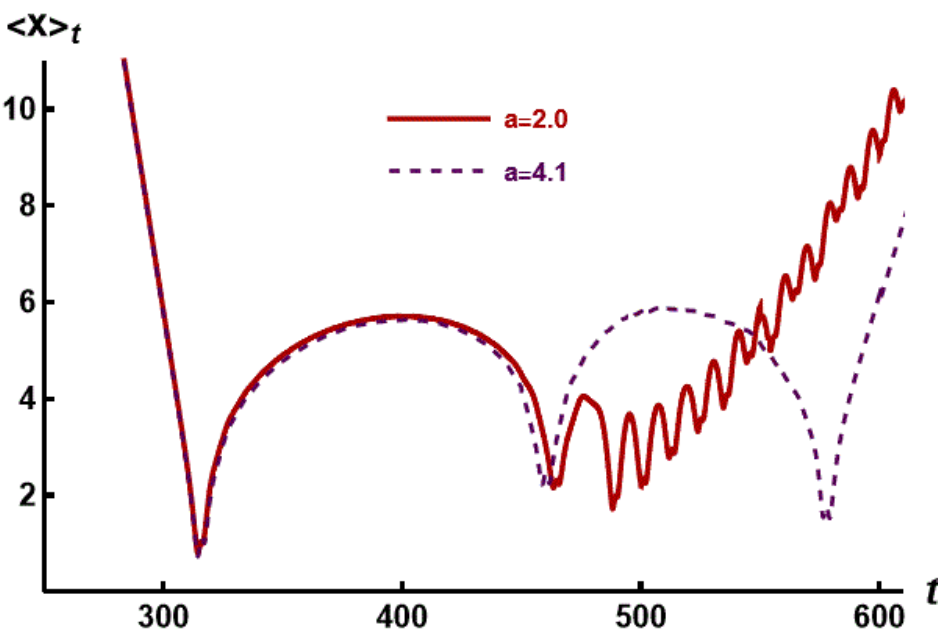}\label{fig:PositionMeanV03157a20a41}}
    \subfigure[ ]{\includegraphics[width=0.45
 \textwidth, height=0.185 \textheight]{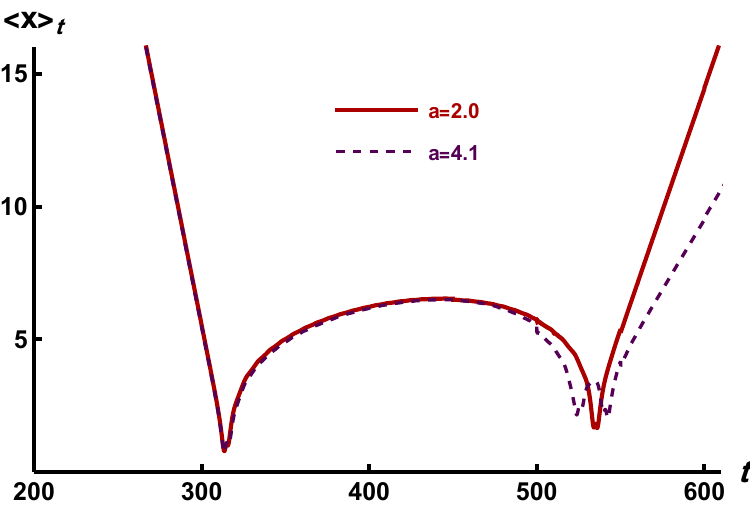}\label{fig:PositionMeanV03166a20a41}}
    \subfigure[ ]{\includegraphics[width=0.45
 \textwidth, height=0.185 \textheight]{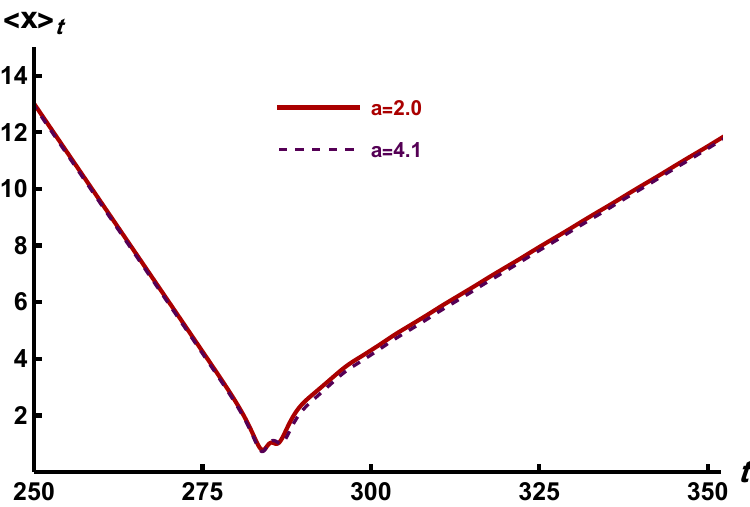}\label{fig:PositionMeanV03500a20a41}}
 \subfigure[ ]{\includegraphics[width=0.45
 \textwidth, height=0.185 \textheight]{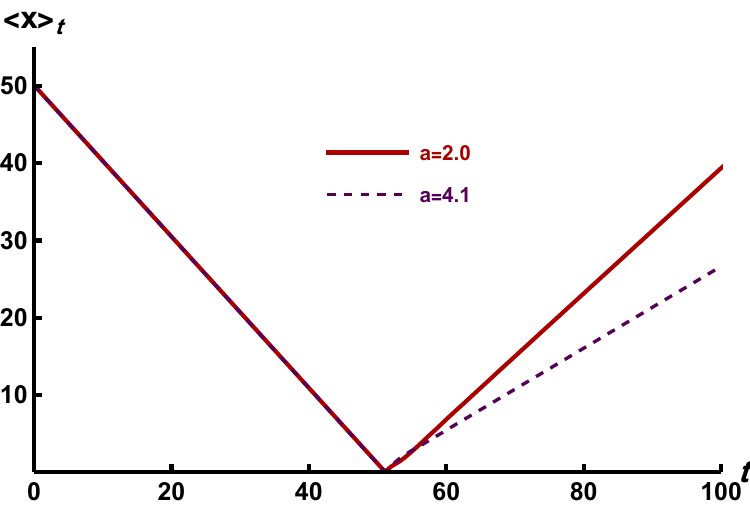}\label{fig:PositionMeanV09800a20a41}}
\\
  \caption{Collision of Kink-antikink pairs within the up sector: Expected position values for varying initial velocities. (a) $v_i=0.3157$, (b) $v_i=0.3166$, (c) $v_i=0.3500$, and (d) $v_i=0.9800$, with an initial position  $X_0=100$.These simulations correspond to parameters $a=2$ and $a=4.1$.}
  \label{fig:PositionMean}
\end{center}
\end{figure*}

Figures \ref{fig:PositionMean} show the time evolution of the collective position of the antikink in the $K\bar{K}$ collision for $a=2.0$, $a=4.1$ and different values of initial velocity. Note that, the kink mass and internal mode frequency are greater for $a=4.1$ as compared with similar quantities for $a=2.0$. At the initial velocity $v_i=0.3157$ with $a=2.0$ (Fig.~\ref{fig:PositionMeanV03157a20a41}) , the kink and anti-kink undergo multiple interactions, while with $a=4.1$ (Fig.~\ref{fig:PositionMeanV03166a20a41}), they scatter after two bounce interactions. The collective analysis, although demonstrates multiple interactions well, but it does not predict some details such as the position of antikink correctly. The clear reason is that, in the case $a=2.0$, initial kink and antikink will melt into each other and create a static bion after the collision. Indeed, the initial anti-kink that followed by the collective position calculations, no longer exists after the collision. For the critical velocity, the obtained results from the collective analysis correctly show that kinks scatter after the first collision, while the final velocity of the scattered particles is greater for $a=2.0$ due to their lower rest mass.

For initial velocities greater than the $v_c$, where the interaction effects with the internal mode are relatively reduced, a great agreement is observed between the collective behavior analysis and the complete numerical simulations (compare Figs.~ \ref{fig:upKakX0100V03166vf01827a20} - \ref{fig:upKakX0100V09800Vf09312a20} with \ref{fig:PositionMeanV03166a20a41}-\ref{fig:PositionMeanV03500a20a41}). But the collective analysis on the (anti)kink position is not able to detect the interesting phenomenon of sector change that occurs for initial velocities greater than the $v_{CS}$ (please see Fig.~\ref{fig:PositionMeanV09800a20a41}).

The ccollective analysis is not able to detect chaotic behavior and fine-tuned interactions between the internal components of kinks, but it predicts the large-scale behavior of collisions and scattering of kinks in a simple way.


\section{Conclusion}
\label{sec:conclusion}
The modified field theory $\Tilde{\varphi}^{(6)}$ has been introduced based on the standard $\varphi^4$ model, containing an adjustable deformation parameter $a$. At limit $a \to \infty$, the deformed model becomes standard $\varphi^6$ field theory, and at the limit $a \to 1$, the model reaches its maximum possible deformation. While standard $\varphi^6$ models do not have an internal mode, the modified model $\Tilde{\varphi}^{(6)}$ has an internal mode inherited from the base model $\varphi^4$. Thus, some behaviors of the $\Tilde{\varphi}^{(6)}$ model are related to the $\varphi^4$ potential and some of its properties are similar to the general dynamics of the $\varphi^6$ field theory. This model has two sectors with asymmetric kink (antikink) solutions, which we have called them "up" and "down" sectors. The down sector kink has greater rest mass than the up sector kink. In the limit $a \to 1$, the mass of the up-sector kink tends to zero while the mass of the kink in the down-sector becomes infinity. At the limit $a \to \infty$, the mass of kinks in both sectors becomes equal to each other. The difference in the kink mass of the two sectors causes very interesting dynamics for the kink collisions in both sectors. According to the initial velocity of colliding $K\bar{K}$ configuration, there exist two different behaviors (in both sectors), which are separated by the critical velocity $V_C$, which is a function of the deformation parameter $a$. 

$K\bar{K}$ configurations in the up sector with an initial velocity lower than the $V_C$, lead to interactions that create different n-bounce windows and finally scatter from each other or bind together in the form of a moving or stationary bion. If the kink (antikink) initial velocity is higher than the critical velocity ($V_C$), the kink and the antikink will move away from each other after a near elastic scattering. If the deformation parameter is large enough, scattered kinks of the up sector (that has a lower rest mass) may appear in the down sector with a greater rest mass, which is rarely observed in other models.

If a $K\bar{K}$ pair of the down sector collide at velocities below the $V_C$, the result is often a bionic bound state. Of course, kinks may move apart after a near-elastic collision. We note that the kink of the down sector (especially for small values of $a$) is massive. Therefore, the high velocity $K\bar{K}$ collision in the down sector, creates a large amount of energy. The result of such collisions will be the emergence of different forms of $K\bar{K}$ pair(s) and/or static or moving bion(s).

Low-speed ($v_i<V_C$) $\bar{K}K$ collisions in the up sector lead to the emergence of the bionic bound states. High-velocity ($v_i>V_C$) interactions are only near-elastic collisions.  One of the unique results of the applied deformation is the possibility of changing the sector of the kinks after the collision. This phenomenon occurs due to the nonlinear dependence of the kink rest mass on the deformation parameter. Such a phenomenon has not been observed in basic models $\varphi^4$ and $\varphi^6$. The $\bar{K}K$ collision in the down sector does not show any critical velocity and the result of the collision is often a set of $K\bar{K}$ pair(s) or bion(s) in the up sector.

 Chaotic behaviors in the final state of the created objects after collisions $K\bar{K}$ and $\bar{K}K$ were investigated. However, the order and overall structure of the resonance windows does not generally change by applying the deformation on the model, the details of the bounce windows and its effects on the produced objects are significant. The collective analysis of (anti)kink temporal position during the collision as well as the dependence of system dynamics and final state of produced objects, on the deformation parameter $a$ and the initial conditions of the interacting kinks were studied. Collective analysis successfully explains the large-scale behavior of interacting kinks. But it cannot determine the details of the interaction, such as the creation of new objects and/or the changing sector of final kink after the collision.

Due to the existence of medium defects and inhomogeneities, an accurate model does not give a true description of a nonlinear system. In practical situations, parts of model parameters are often changed and we need to apply some deformation to the standard field theories. The nature of the deformed model gives a unique combination of the properties inherited from the basic models and the deformation mechanism. Such deformed field theories provide better approximations for the behavior of the system under experimental conditions. The presented method is applicable to many realistic models. In further works, more complicated deformed models can be investigated.




\end{document}